# Effect of Accelerated Thermal Degradation of Poly(Vinyl Chloride): The Case of Unplasticized PVC

Marwa Saad[1], Marek Bucki[2], Sonia Bujok[3], Dominika Pawcenis[1], Tjaša Rijavec[4], Karol Górecki[1], Łukasz Bratasz[3], Irena Kralj Cigić[4], Matija Strlič[4], Krzysztof Kruczała[1*]

[1] Faculty of Chemistry, Jagiellonian University in Krakow, Gronostajowa 2, 30-387 Krakow, Poland

[2] Jagiellonian University, Doctoral School of Exact and Natural Sciences, Gronostajowa 2, 30-387, Krakow, Poland

[3] Jerzy Haber Institute of Catalysis and Surface Chemistry, Polish Academy of Sciences, Niezapominajek 8, 30–239 Kraków, Poland

[4] Heritage Science Laboratory Ljubljana, Faculty of Chemistry and Chemical Technology, University of Ljubljana, Večna pot 113, 1000 Ljubljana, Slovenia

E-mail: kruczala@chemia.uj.edu.pl

**Abstract**

The thermal degradation of unplasticized poly(vinyl chloride), PVC, was comprehensively investigated through the application of spectroscopic techniques, as well as contact angle measurements (CA), dynamic mechanical analysis (DMA), and size-exclusion chromatography (SEC). To study the effect of relative humidity (RH) on the deterioration of unplasticized PVC, two regimes of accelerated degradation experiments were selected: low RH (max. 30% RH) and high RH = 60% levels, which corresponds to usually the highest RH in heritage institutions equipped with an HVAC system. Nuclear magnetic resonance (NMR) and infrared spectroscopy (FTIR) did not reveal any significant changes in the material during its degradation up to 20 weeks at temperatures ranging from 60°C to 80°C. Notable changes were observed in the Raman and UV-Vis spectra, indicative of the formation of conjugated carbon-carbon double bonds. The formation of polyenes was responsible for the yellowing of samples. Notwithstanding, the aforementioned changes did not lead to a notable decline in the mechanical properties, as evidenced by DMA and SEC measurements. EPR measurements demonstrated the formation of radicals at 60°C, and in the sample degraded at 80°C the presence of radicals was evident. This indicates that a radical degradation mechanism cannot be excluded even at such low temperatures.

**Keywords:** PVC thermal degradation, heritage objects, unplasticized PVC, spectroscopy, DMA

**1. Introduction**

The extensive development of polymeric materials, where the main ingredients are synthetic polymers, is one of the twentieth century's biggest technological and industrial achievements. With a huge variety of forms, colors, and textures, the ease of molding, edge-banding, and modifying, commercial polymeric materials also found their way into artists' studios. They have improved artistic expression in architecture, design, and artistry, enhancing modern art's capacity for interpretation and communication. Plastics of the twentieth century challenged their conservators with the problem of the degradation of polymeric materials and the resulting loss of objects' value. The threat of irreversible loss of the tangible cultural heritage of the $20^{th}$ and $21^{st}$ centuries is a result of industrially manufactured plastics being intended for short-term use. They have in-built lifetimes ranging roughly from 1 year (supermarket carrier bag) to 50 years (u-PVC window frames), and, when exposed to unsuitable environments, they may exhibit rapid degradation, leading to the loss of their original properties. Another problem is related to the lack of control over material qualities, properties, and quantities resulting from modifications done by artists, which may have unexpected detrimental effects on object durability [1].

The deterioration of artifacts made from plastics was first reported in the literature in the late 1980s [2]. Estimates suggest between 15–30% of plastic objects in institutions across Europe are in poor or unacceptable condition for display [3]. The deterioration of polyvinyl chloride (PVC) can be accelerated as a result of erroneous formulations. This phenomenon may give rise to changes that were not foreseen by either the artists or the collectors of these artworks. Yellowing or darkening of plastic works of art during exhibition and storage is recently one of the main issues in the case of objects made of PVC [4].

PVC is one of the "big five" large-tonnage plastics today. More than 5 million tons were manufactured in Europe in 2018, representing 10% of overall plastic production [5], and over 50 million tons were produced globally in 2022, amassing nearly 13% of all polymers [6]. A wide-ranging survey of museum collections in France, the Netherlands, and the UK was carried out in the scope of the POPART project [7]. As a result, it was found that PVC was present in all collections and represented around 13% of all plastic objects. This study revealed that 68% of objects were in a good or fair state, 25% were in a state of significant decay and 7% were severely damaged. Cellulose acetate (CA), cellulose nitrate (CN), and PVC represented 40% of all objects in poor state [7].

Reporting of such problems has been prompted by museums worldwide. Keneghan et al. [3] published a survey of 7900 plastic objects from the Victoria and Albert Museum's collection in London (England). 10% of surveyed objects suffered from degradation and showed evidence of chemical damage (classified as bloom, brittleness, or discoloration). Klisińska-Kopacz et al. [8] reported chemical damage to artworks made of PVC from the National Museum in Krakow (Poland).

PVC can be produced in two formulations – plasticized and unplasticized. Unplasticized PVC is hard and rigid, examples of which are PVC objects from the Postcards 1968-1974 collection in Harvard Art Museums [4], while, by contrast, plasticized PVC is softer and more flexible, hence used more commonly in toys, design items and sculpture arts. PVC objects are degraded through two processes: degradation of PVC polymer chain and plasticizer loss [9,10]. PVC polymer degrades through dehydrochlorination, in which the polymer is converted to a hydrocarbon with a conjugated polyene system, releasing hydrochloric acid and accelerating the degradation process [11]. This conjugated polyene structure absorbs progressively longer wavelengths of light as the process continues, causing an increasing yellowing of PVC plastics over time [12]. Additionally, Shashoua et al. [13] reported cross-linking and chain scission reactions occurring during PVC degradation which alter the physical properties and lead to embrittlement. Plasticizer loss has been reviewed by many scientists (King et al. [10], Rijavec et al. [9], and Ledoux [4]). According to King et al. [10], plasticizer migration initially takes place through evaporation or absorption by a secondary material, such as a storage box, which further encourages the migration of more plasticizers to the surface due to a concentration gradient. Shashoua et al. [13] examined plasticizer loss by accelerated degradation of model plasticized PVC for 65 days and reported that the main source of the loss was due to plasticizer diffusion toward the surface.

The degradation of PVC with other different additives was investigated by many researchers [14–17]. Gupper et al. [18] compared the degradation of unstabilized PVC with the degradation of PVC stabilized with calcium stearate, zinc stearate, and zinc chloride, which are common heat stabilizers used to avoid the decomposition of PVC during melt processing. Arkis et al. [19] observed the effect of organotin as a stabilizer over PVC thermal degradation at 140°C, 160°C and 180°C and proved that it is a good heat stabilizer for plasticized PVC while not having any detrimental effects on mechanical properties [19]. On the other hand, according to Bacaloglu et al. [20,21], the presence of zinc or tin carboxylates can be responsible for a very fast discoloration of PVC – Zn or Sn carboxylates act as stabilizers until the corresponding halides are formed which marks the start of fast degradation due to the formation of Lewis acid sites by alkyl halides exhibiting a catalytic effect for further dehydrochlorination reaction.

Most literature related to the thermal degradation of rigid unplasticized PVC during artificially accelerated tests deals with the degradation mechanisms at high temperatures (120-200°C range) [22–26], i.e. significantly different than typical environmental conditions encountered in heritage institutions. Less attention has been paid to artificial thermal degradation at moderate temperatures and high relative humidity (RH) that could be more representative of the conditions of the museum environment. Therefore, in this study we focused on unplasticized PVC stabilized with organotin (denoted PVC1) which was artificially degraded at lower temperatures ranging from 60°C to 80°C with a wide RH range of up to 60%. Analysis of the degraded samples was done by Raman spectroscopy, ATR-FTIR, spectrophotometric evaluation of color changes, UV-Vis, SEC, contact angle measurements, XPS, NMR, EPR and dynamic mechanical analysis (DMA) to recognize the fundamental link between chemical degradation and changes of macroscopic mechanical properties.

## 2. Experimental:
## 2.1. Materials

Rigid transparent PVC foil of 0.2 mm thickness (produced in 2019 by Alfatherm s.p.a., herein denoted PVC1) was used to investigate the thermal degradation of unplasticized PVC. PVC1 is stabilized mainly with Irganox 1076 (octadecyl-3,5-bis(1,1-dimethyl ethyl)-4-hydroxy benzene propanoate – an ester of stearyl alcohol and propionic acid modified with large groups with significant steric hindrance providing thermo-oxidative stability to PVC).

## 2.2. Accelerated degradation

For accelerated degradation, the PVC1 samples with the dimensions of 2x2 cm were placed in a climate chamber (BINDER Deutschland GmbH) at 60°C and 80°C with relative humidity control (RH = 60%) for over 20 weeks. In order to investigate the humidity effect, another sample sets were degraded under low humidity regimes: at 60°C with RH = 20%, at 70°C with RH = 30%, and at 80°C without controlled humidity (using the Mollier diagram, the RH was no higher than 7%) for up to 22 weeks. The samples were also degraded in a dryer at high temperatures (120°C and 140°C) for up to five days to differentiate between degradation mechanisms at low and high temperatures.

**2.3 Methods**

Raman spectra were measured using a Renishaw inVia Raman microscope 2000 system with the following parameters: an excitation wavelength of 514 nm, an acquisition time of 60 s, a laser power of 5% and spectral resolution of 4 cm$^{-1}$, and a spectral range of 250–3500 cm$^{-1}$. Each spectrum was the average of five scans obtained from different spots on the sample surface. Savitzky–Golay smoothing filter was used to minimize noise.

The fundamental vibrations and associated rotation-vibration structures were studied in the mid-infrared (IR) region, 4000-500 cm$^{-1}$, by FTIR Nicolet 6700 Thermo Scientific spectrometer equipped with an ATR device. The ATR-FTIR measurements were performed on both sides of the samples at three different places with a data interval of 0,482 cm$^{-1}$, 32 scans were averaged for each measurement, and a Happ-Genzel apodization and Mertz phase correction were used.

The X-ray photoelectron (XPS) spectroscopy technique examined the surface compositions of degraded and undegraded PVC samples (R3000 Data Scienta Analyzer-Prevac). The base pressure in a vacuum chamber was below $5\times10^{-9}$ mBar, and a monochromatized Al-Kα source with a 250 W at 1486.6 eV emission energy was used. The scale of the binding energy value was adjusted to the C 1s reference peak at 284.8 eV. The composition and chemical state of samples surface was analyzed in terms of areas and binding energies of O 1s, C 1s, and Cl 2p photoelectron peaks. The spectra were fitted using CasaXPS software Version 2.3.24PR1.0 (Casa Software Ltd., Teignmouth, UK) [27]. The XPS peak fitting was performed using the Shirley background and the GL(30) function. Maximum half-peak widths were limited to 2.0 eV in order to avoid excessive overlapping of the peaks and to assign reliable binding energies.

Water contact angle (WCA) measurements were performed at room temperature in the air. Using the drop shape analysis-profile device with a tiltable plane (DSA-P, Krüss, Germany), advancing contact angles of water were measured. Initially, drops of Milli-Q Ultrapure Water, 15 µL in volume, were applied on the surface of the PVC1 sample using a variable-volume micropipette. After that, the sample that was supporting the drop was tilted at a steady rate (1°s$^{-1}$), and photos of the drop were concurrently captured. The hydrophobicity/hydrophilicity of a PVC1 surface can be estimated by contact angle measurement. Contact angle, $\theta$, is a quantitative measure of the wetting of a solid by a liquid. By drawing a tangent line from the drop, between the tangent line and the solid surface is defined geometrically the contact angle formed by a liquid at the three-phase boundary where liquid, gas, and solid intersect, as described by Young-Dupré [28]. The values that were given are the average of three measurements made at various locations on the surface of PVC1 samples.

Color changes and yellowing index (a calculated number used to describe a change in color from a clear or white sample toward yellow) were determined with the color analyzer ColorQuest XE using a color space colorimeter equipped with standard illuminant C, and the CIELab parameters, $L^*$, $a^*$ and $b^*$, were determined. The vertical axis $L^*$ is a measure of lightness, from black (0) to white (100), while on the hue-circle, $a^*$ is a measure of redness (or $-a^*$ of greenness) and $b^*$ of yellowness (or $-b^*$ of blueness). Hue angle (H*) and chroma (C*) values are obtained from $L^*$, $a^*$ and $b^*$ coordinates [29].

The electron paramagnetic resonance (EPR) measurements were performed with Bruker ELEXSYS 500 and 580 spectrometers. The spectra were acquired and post-processed using the Xepr data system. The ELEXSYS 500 was equipped with the X-band super high-sensitivity cavity ER 4122 SHQE, and spectra were recorded with a 100 kHz magnetic field modulation. In standard experiments, the microwave power was 2 mW, sweep width 20 mT, modulation amplitude 0.2 mT, and 4 scans were acquired. The ELEXSYS 580 was equipped with the Q-band SHQE1021 cavity. Spectra were recorded with a 50 kHz magnetic field modulation, 2 mW microwave power, 0.7 T sweep width, 0.3 mT modulation amplitude, and 1 scan was acquired. All EPR spectra were recorded at room temperature.

UV-Vis electron absorption spectroscopy was used to examine the changes in the electronic structure of degraded PVC1 samples, collected on an Agilent Technologies Cary 7000 UV-Vis-NIR spectrometer. Measurements were performed in the 200 to 800 nm wavelength range, with a 1 nm data interval and a 600 nm min$^{-1}$ scan rate (0.1 s averaging time). All samples were measured repeatedly, resulting in a few, consecutive scans, from which the best ones were chosen. For derivative calculations, a 20-point smoothing of data was used by the Lowess method.

The elemental composition of the samples was determined by elemental analysis (Vario MicroCube). The method permitted the quantification of the carbon, hydrogen, nitrogen, and sulfur contents in the samples.

The nuclear magnetic resonance (NMR) measurements were performed using a Jeol 400 MHz YH spectrometer. Prior to the measurements, samples were dissolved in deuterated THF using sonication, resulting in approx. 3 mg mL$^{-1}$ concentration after passing through a 0.45 µm PTFE syringe filter. For the $^1$H NMR measurements, 16 scans were collected and averaged to create the final spectrum, whereas for the $^{13}$C NMR measurements, 256 scans were collected.

The dynamic mechanical analysis (DMA) was performed using the Triton Tritec 2000 DMA Thermoanalyzer. Time scans were run at 25°C at the 1 Hz frequency at a constant deformation of 0.002 mm (ca. 0.02% strain) in tension mode. For the comparison of sample stiffness after a certain period of degradation, values of the storage moduli $E'$ at 1 Hz were determined as the average of 2-3 measurements. Additionally, a temperature scan in the range of 30-90°C with a heating rate of 2°C min$^{-1}$ was performed to determine glass transition temperature $T_g$. Specimens of dimensions ca. 0.2x10x20 mm$^3$ were used for all experiments.

Molar mass distributions were determined using the size exclusion chromatography (SEC-MALLS-DRI) technique. The methodology of samples' preparation and conditions of chromatographic analyses were as described previously [30]. Briefly, PVC1 samples were dissolved in tetrahydrofuran (HPLC purity, Chemland) at a concentration of 1 mg mL$^{-1}$ and left overnight for complete dissolution. The next day, each solution was put in an oil bath thermostated at 55°C for 5 h in order to break aggregates, that could be formed during dissolution. Prior to injection, samples were filtered using syringe filters (PTFE, 0.45 µm, Biospace, Poland). The chromatographic system consisted of a Water Breeze system: isocratic pump Waters 1515, autosampler 717+, column oven (thermostated at 30°C). Tetrahydrofuran was used as a mobile phase at a flow rate of 1 mL min$^{-1}$. Separation was achieved in two mixed-bed DVB Jordi columns (25 × 1 cm, separation range 100–10,000,000 g mol$^{-1}$), equipped with a Jordi precolumn. The eluting fractions were directed to a multiple angle laser light scattering (MALLS) detector (Dawn Heleos, Wyatt Technology) and differential refractive index detector (DRI) (Optilab t-rEX, Wyatt Technology, thermostated at 30°C). Molar masses were calculated according to the Rayleigh equation and dn/dc equal to 0.100 g mL$^{-1}$.

Microwave digestion was carried out using Milestone's ETHOS UP system. The manufacturer's application note method [31] was adapted as follows: a small amount of sample (25.000 mg) was accurately weighed, 8 mL of concentrated $HNO_3$ and 3 mL of concentrated $H_2SO_4$ were added. The vessels were heated to 210°C in 25 min at 1800 W, maintained for 15 min before cooling down. The solution was then transferred to a 50 mL flask and diluted with 1% $HNO_3$ prepared in MQ water. The solutions were filtered through a 0.45 μm nylon filter. The concentration of metal ions in **Table S1** was determined using Agilent 5100 ICP-OES. Five replicates of the sample were analyzed. The coefficient of variation of the method is 1%. Inorganic Ventures VAR-CAL 1 standard (100 mg $L^{-1}$: Ti, Sb, Mo, Sn) was used to prepare mixed-metal calibration standards in the range of 1 - 1000 μg $L^{-1}$ in MQ water (>18 MΩcm). Additionally, the values were then cross-checked using Perkin Elmer NexION ICP-MS after microwave-assisted mineralization of PVC1 samples.

To evaluate the effect of storing rigid PVC objects with other items, especially in the context of heritage institutions, regarding the release of volatile compounds, the standard Oddy test was conducted on PVC1 samples, adapted from the 20190226_OT_1 protocol (Buscarino, Stephens, and Breitung 2021) of the Metropolitan Museum of Art's Department of Scientific Research. Briefly, 2 g of undegraded PVC1 was placed in a glass bottle with a stopper sealed using PTFE tape with four vials – one filled with water to provide a 100% RH, the other ones with metal foils of copper, lead and silver. Two such kits were prepared with the PVC1 foils, and two without the sample as a reference. The glass bottles were placed in a dryer set at 60°C for 28 days. After the time period had elapsed, the metal foils were investigated for any signs of corrosion visually with the naked eye and under a Leica DM 2000 microscope equipped with a revolver of objectives, from which the 10x and 20x were used.

## 3. Results and Discussion

The unplasticized PVC was used to separate the effect of thermal degradation on PVC mechanical properties from the changes due to plasticizer migration and its influence on the degradation processes. A previous investigation by Rijavec et al. [32] revealed that the rate of discoloration is affected by the content of plasticizers in the way that higher plasticizer content leads to a slower yellowing rate which is in agreement with the findings of Shashoua et al. [33] who reported that the presence of plasticizers inhibited the degradation of PVC. Therefore, unplasticized PVC represents the worst-case due to the highest possible concentration of chlorine and the expected highest degradation rate [32].

To the best of our knowledge, only a few articles on the degradation processes of unplasticized PVC at low temperatures below $T_g$ have been published so far [30,32,34,35]. Therefore, firstly the $T_g$ of the PVC1 was evaluated using dynamic mechanical thermal analysis (DMTA) to select appropriate degradation conditions. DMTA scans indicated a $T_g$ value of 81.8°C as the average of the two tan$\delta$ peaks maxima (**Figure S1**) [36,37]. Based on this, the maximum degradation temperature for this study was selected as 80°C.

To confirm the composition of PVC1, the content of plasticizers was investigated according to the procedure described in [38] using gas chromatography with mass spectrometry detection. The content of plasticizers in PVC1 was below the detection limit (<3%). Furthermore, as will be described later in the spectroscopic section, FTIR and Raman spectra (**Figure 4** and **5**) do not contain peaks originating from typical plasticizers, which allows us to confirm that PVC1 is unplasticized. The investigated PVC1 samples contained 0.10%$_{at}$ tin from the stabilizer, as determined by ICP-OES and ICP-MS after complete microwave digestion of organic compounds (**Table S1**).

Although the moisture sorption isotherm for PVC1 at 18°C demonstrates that the material absorbs only up to 0.09% water when the relative humidity increases from 10% to 80% (**Figure 1**), even this small increase in the water content can affect the polarity of the medium, thereby accelerating the dehydrochlorination reaction. Specifically, PVC1 was exposed to accelerated degradation at 70°C with 30% and 80% RH. The rate of dehydrochlorination, quantified as the degree of yellowing, increased from 0.02 $\Delta b^*$ per day at 30% RH to 0.14 $\Delta b^*$ per day at 80% RH. These results indicate that relative humidity plays an important role in the degradation process [32].

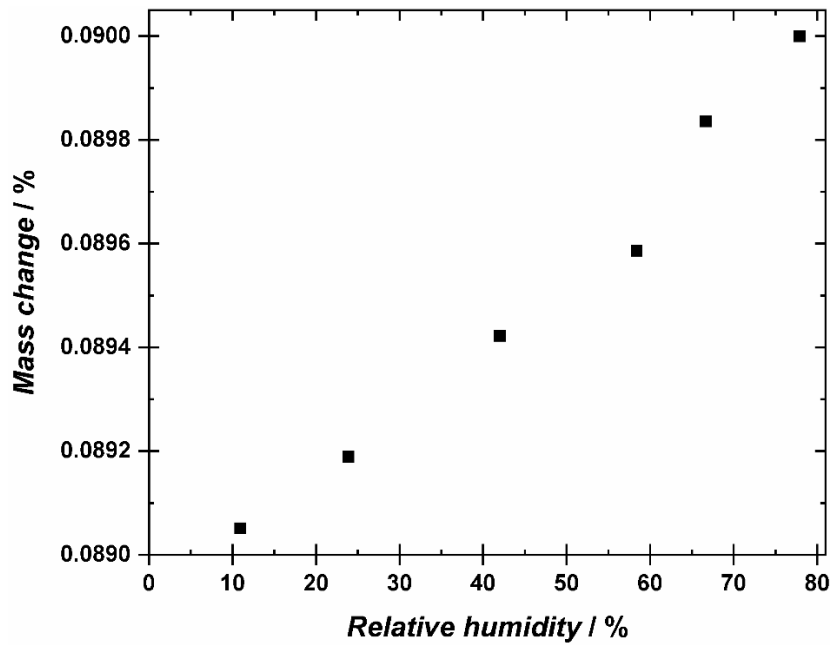

**Figure 1.** Moisture sorption isotherm for PVC1 performed at 18°C in the relative humidity range from 10% to 80%.

### 3.1. Color changes

Firstly, the degradation of PVC1 samples at various relative humidity conditions:
- Case I - low relative humidity (up to 30% RH),
- Case II - higher relative humidity of 60% reflecting typical maximum RH level at controlled (air-conditioned) museum conditions

was examined to investigate the effect of the RH by comparing the color changes of the undegraded and degraded samples, which are potentially noticeable to the naked eye, and thus of high significance to museum objects.

Visual inspection of PVC1 samples degraded at 60°C, RH = 20% and 70°C, RH = 30% (Case I) revealed minor color changes visible to the human eye (after 7 weeks of degradation at 70°C, RH = 30%). Increasing the temperature to 80°C without RH control (RH below 10%) results in a significant acceleration of the discoloration as samples become visibly yellow much faster (already after 6 weeks of degradation) and color changes are much more significant than at 60°C (20% RH) and 70°C (30% RH) degradation conditions. In Case II, corresponding to 60% RH, minor color changes at 60°C are observed by the naked eye in the entire degradation time, whereas at 80°C, extensive discoloration starts from 4 weeks of the experiment. Additionally, all samples remain visually transparent.

Color measurements were performed to quantify these observations. Obtained results are shown in **Figure 2** and **Table S2** summarizing the evolution of $\Delta E$ and single coordinates values ($\Delta L$, $\Delta a$, and $\Delta b$) as a function of the experimental time. Two $\Delta E$ levels are distinguished as the so-called just noticeable color change at $\Delta E$ usually larger than 1.8 [39] (only experienced observers may notice difference defined by $\Delta E$ from 1 to 2) and a different color when $\Delta E$ exceeds the value of 5 [39]. CIEDE2000 color-difference formula was used in the computation of the color change. Given a pair of color values in CIELab space, we denote the CIEDE2000 color difference between them as follows: $\Delta E_{00} = \Delta E_{00}(L^*_1, a^*_1, b^*_1; L^*_2, a^*_2, b^*_2)$. The implementation of the $\Delta E_{00}$ function in several steps is described by Sharma et al. [40].

As shown in **Figure 2** and **Table S2,** there are no differences observed in the degraded and non-degraded samples at 60°C either at low RH (Case I) or RH = 60% (Case II) in the lightness $\Delta L$ and the parameter $\Delta a$, whereas the main change is the yellowing characterized by the evolution of the parameter $\Delta b$ as it increases with the degradation time. The $\Delta b$ reaches 0.12 for Case I and 1.90 for Case II at 60°C for 20 weeks of degradation.

For the samples degraded at 80°C either Case I or Case II, the results in **Figure 2c, d** and **Table S2** show that the change in PVC1 color is a combination of lightness ($\Delta L$), redness ($\Delta a$) and yellowness ($\Delta b$) components. The $\Delta E$ for samples degraded at 80°C is higher at RH = 60% (Case II) than those degraded at low RH (Case I), which indicates that humidity accelerates the degradation. A similar observation was previously reported by Rijavec et al. [32], who observed that a higher humidity accelerates the yellowing of PVC as an increase in the storage RH from 20% to 70% at 25°C for the same sample led to a decrease of the predicted lifetime from 36 years to 27 years.

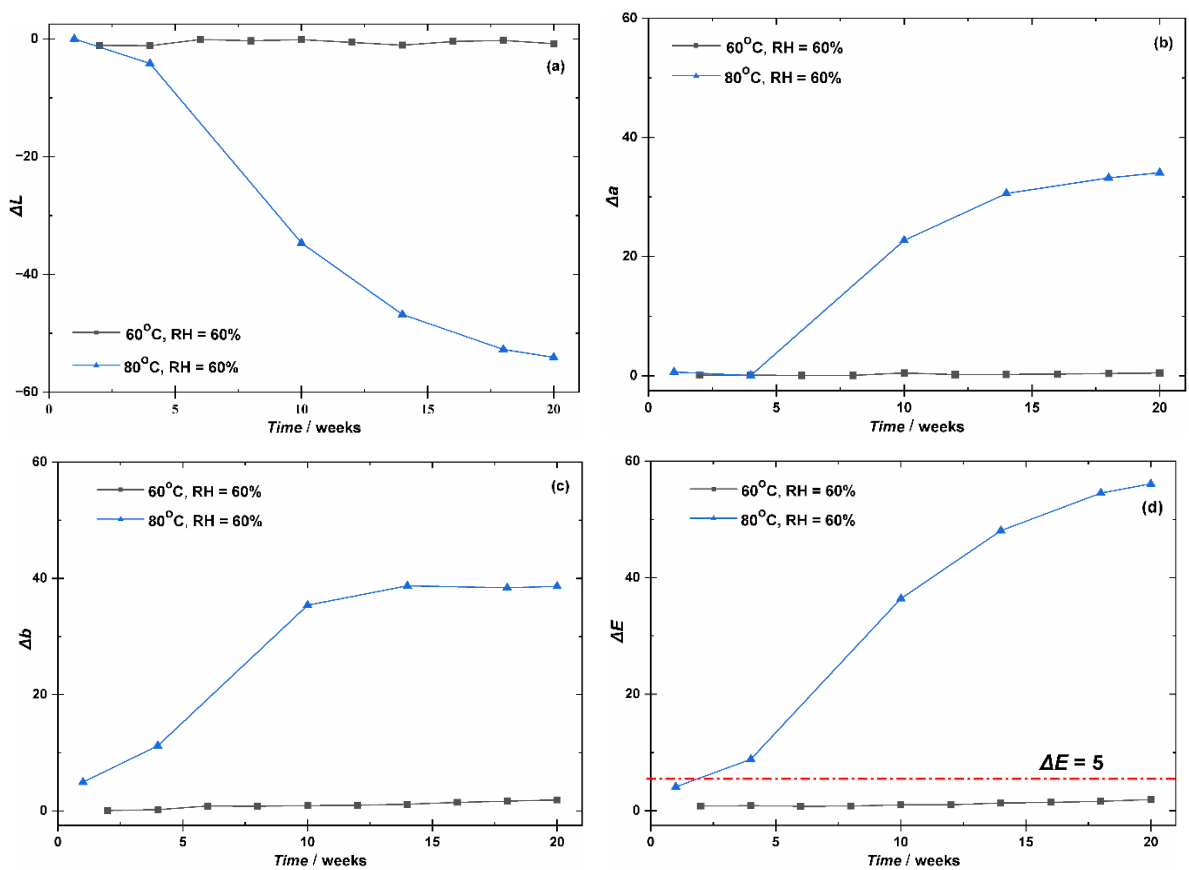

**Figure 2.** Color coordinates of the degraded PVC1 in the condition indicated in the figure (Case II) by CIELab color space: *ΔL* **(a)**, *Δa* **(b)**, *Δb* **(c)**, and *ΔE* **(d)** calculated with reference to undegraded PVC1.

## 3.2. Spectroscopic analyses: mechanism of PVC degradation below $T_g$
### 3.2.1. Electron paramagnetic resonance (EPR) spectroscopy

The degradation of PVC may occur via a variety of mechanisms, such as free-radical [41], ion-pair/quasi-ionic [42], ion-radical (polaron) [33], and unimolecular mechanisms [42]. The degradation mechanism of PVC is dependent upon the specific conditions under which the PVC is degraded, and the additives present in the material. In order to investigate the variation between PVC mechanisms as a function of time and temperature, PVC1 samples were subjected to degradation at relatively high temperatures (120°C and 140°C) for 3 and 5 days, respectively, and subsequently investigated by EPR which provides information on the presence of free radicals. The non-degraded sample exhibits a broad signal **S1** with a *g* value of approximately 2.03, whereas, after 3 days of degradation at both elevated temperatures, visible peaks **S2** appeared at
12

$g$~2.003, and the intensity increased with degradation time, as shown in **Figure S2**. It can be reasonably assumed that the signal **S2** is associated with the formation of carbon radicals within the polymer chain in the process of degradation [43]. Concerning signal **S1**, it may be hypothesized that its origin lies in iron impurities detected by ICP-MS (**Table S1**). This assumption is further supported by the appearance of a signal in the $g$~3.9 region (**Figure S3**), a typical value for iron ions. Nevertheless, the precise origin of the signal **S1** is yet to be determined. However, this does not constitute the primary concern of this study.

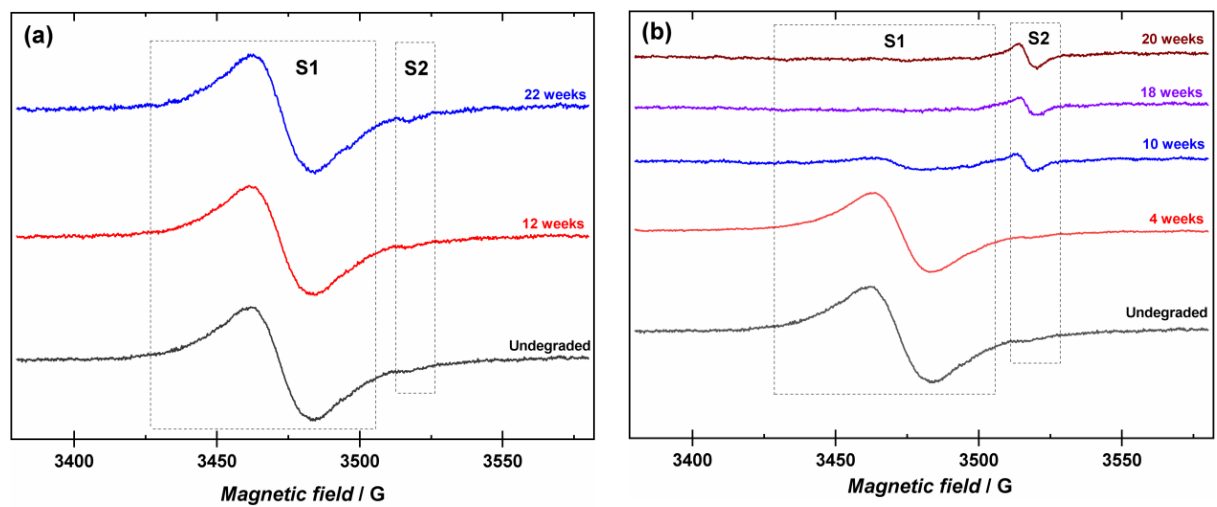

**Figure 3.** X-band EPR spectra of undegraded and PVC1 degraded at 60°C **(a)** and 80°C with RH = 60% **(b)** for the indicated time (**Case II**).

EPR spectra of PVC1 thermally degraded at 60°C for up to 22 weeks and at 80°C for up to 20 weeks are shown in **Figure 3**. The broad **S1** signal becomes very weak after degrading the PVC1 sample for ten weeks at 80°C and 60% RH and completely disappears after 18 weeks of degradation at 80°C. The signal **S2** appeared also after degradation at 60°C and 80°C for a prolonged time and its intensity is correlated with the amount of stable radical contents in the studied sample. Radical formation started at 60°C, RH = 60% after 12 weeks of degradation, confirmed by a small peak at $g = 2.003$. Moreover, the peak becomes more intense by raising the temperature to 80°C at the same humidity, indicating that more radicals are formed at higher RH. According to many researchers [21,23,44–49], the radical mechanism (**Scheme 1**) occurs in PVC degradation at temperatures higher than 120°C, and at lower temperatures, the ionic mechanism prevails. However, according to Starnes [42] review, no evidence exists that the free radical mechanism cannot occur at low temperatures.

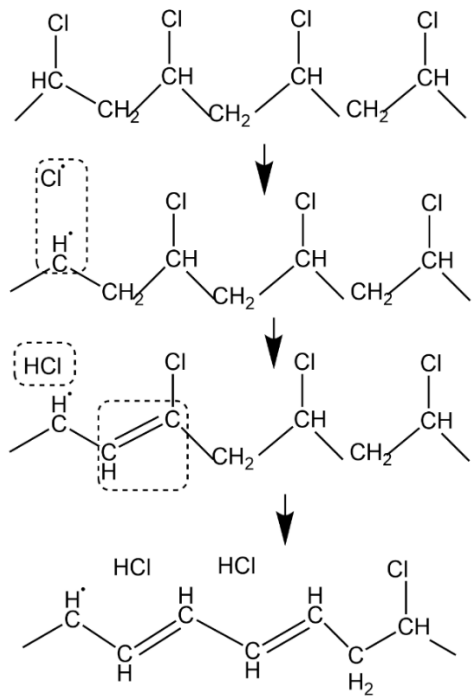

**Scheme 1.** The free radical mechanism.

Stranes [42] also pointed out that two research groups [50,51] proposed a 151 kJ/mol value for the energy input required for the radical mechanism to start. The presence of a signal from radicals in the EPR spectra indicates that the radical mechanism of PVC degradation cannot be excluded, even at lower degradation temperatures.

**3.2.2. Infrared (IR) and nuclear magnetic resonance (NMR) spectroscopy**

The functional groups of the PVC were analyzed by ATR-FTIR spectroscopy, and the spectra of PVC1 degraded at 60°C and 80°C at RH = 60% (Case II) are shown in **Figure 4a** and **Figure 4b**, respectively. The absorption bands visible at 610, 635 and 690 cm$^{-1}$ of the PVC1 in the spectrum correspond to the different stretching vibration modes of the C-Cl bond, whereas those at 1088 and 1095 cm$^{-1}$ belong to the stretching vibration of the C-C bonds. The vibration at 1197 cm$^{-1}$ corresponds to the out-of-plane bending vibration of the C-H bonds, while those at 1253 and 1329 cm$^{-1}$ correspond to the in-plane and out-of-plane C-H vibrations in the CH-Cl groups. The CH$_2$ group vibrations appear at 955 and 1425 cm$^{-1}$. The other vibrations appearing between 2850 and 3010 cm$^{-1}$ belong to the different vibration modes of the CH and CH$_2$ bonds [52–54]. The maximum at 1730 cm$^{–1}$ in the ATR-FTIR spectrum is attributed to the C=O groups and may be related to the organotin stabilizer which contains carbonyl and carboxylic groups [19,55]. No significant difference was observed in ATR-FTIR with increasing degradation time for both temperatures (60°C and 80°C). Tabb and Koenig [56] observed similar findings during their study of plasticized and unplasticized PVC by FTIR. They noticed that the change in the intensity and shape of the spectra were visible only in the plasticized PVC. Similarly, no significant difference was observed in ATR-FTIR for PVC1 degraded at lower humidity as presented in **Figure S4**.

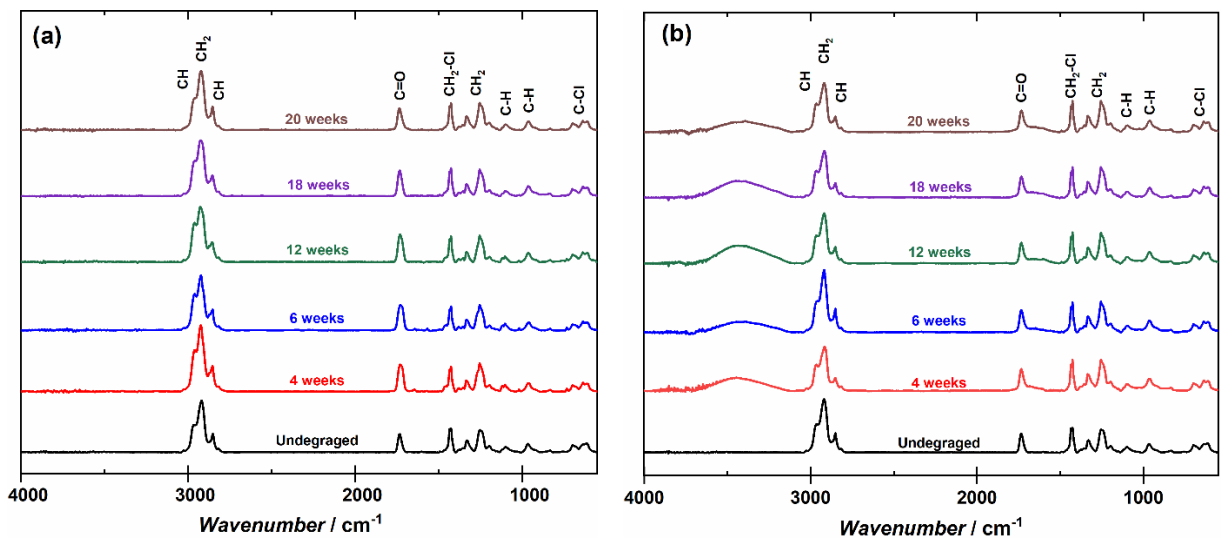

**Figure 4.** ATR-FTIR for PVC1 degraded at 60°C, RH = 60% **(a)** and 80°C, RH = 60% **(b)** referring to Case II for the time indicated in the figure.

NMR spectra of both undegraded and degraded samples were found to be nearly identical, indicating a low concentration of newly created moieties due to dehydrochlorination upon degradation. Differentiation of signals in the $^1$H NMR (**Figure S5a**) results from protons being located at a non-identical distance from strongly electronegative chlorine atoms, which causes deshielding and an increase in the chemical shift. Most strongly shifted, at around 3.6 ppm, are hydrogen atoms connected with the carbon atoms with chlorine. The next signal, at 2.4 ppm, is created by protons located in the vicinity of a chlorine atom, on adjacent carbon atoms. The last one, located around 1.7 ppm, comes from hydrogen atoms the furthest from the chlorine atoms. On the $^{13}$C NMR (**Figure S5b**), there are only two signals present – one upfield at around 24 ppm from carbon atoms without a chlorine atom attached and one downfield at around 66 ppm from carbon atoms with a chlorine atom.

### 3.2.3. Raman and UV-Vis spectroscopy

Raman spectroscopy has been pointed out to be one of the most suitable tools to investigate conjugated bond systems, being extensively used in the characterization of the degradation of PVC [8,18,57–59]. **Figure S6** gives the typical Raman spectra with excitation wavelength $\lambda = 785.0$ nm of PVC1 artificially degraded at 80°C up to 22 weeks. The 637 and 695 cm$^{-1}$ peaks were attributed to C-Cl stretching vibrations, making the polymer identification possible [60]. The resonance at about 1430 cm$^{-1}$ is attributed to $\delta$ (CH$_2$) [8]. Small weak bands at 997-1053 cm$^{-1}$ were attributed to the COO$^-$ group of the stabilizer (organotin or Irganox 1076) [19,55]. The attribution is corroborated by ATR-FTIR spectra, wherein the peak at 1730 cm$^{-1}$ indicative of carbonyl groups, is clearly visible (**Figure 4**). We did not observe polyene bands at the excitation laser line $\lambda = 785.0$ nm, not only because this excitation corresponds to a non-resonance condition but also due to the very low contents of the polyene in our sample. This latter conclusion is consistent with the mild degradation temperature. However, since earlier studies show that polyene bands can be detected in the spectra of PVC even at non-resonance excitations ($\lambda = 785.0$ or 1064.0 nm) [61,62], we assume that the Raman bands of long polyenes under red laser at excitation of $\lambda = 785.0$ nm can be detected in PVC spectra only for high polyene content. Therefore, another excitation wavelength, 514.0 nm (green laser), was used to investigate the degradation of PVC.

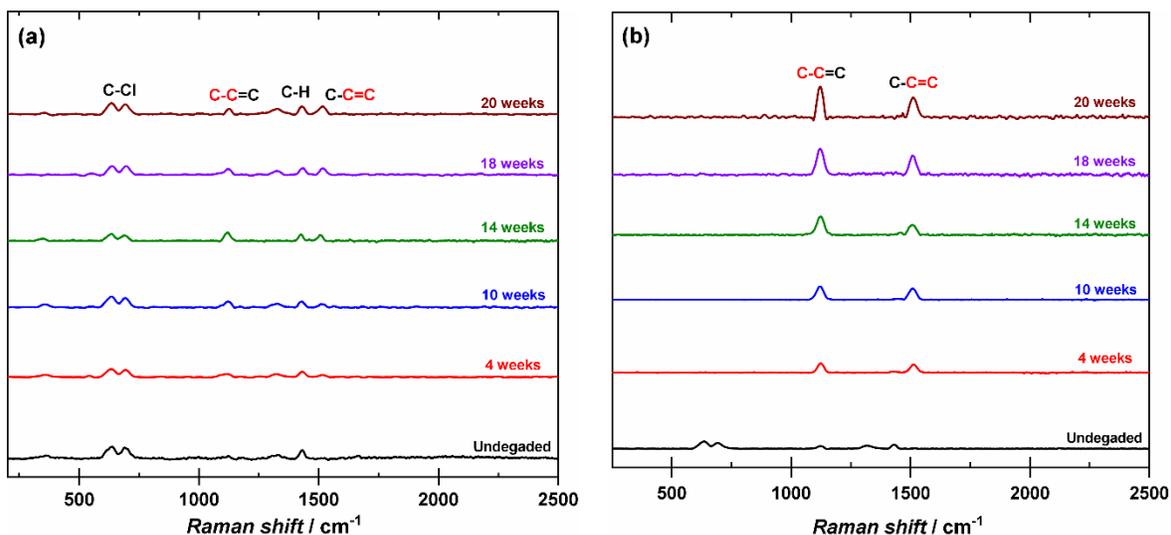

**Figure 5.** Raman spectra of PVC1 degraded at 60°C, RH = 60% **(a)** and 80°C, RH = 60% **(b)** referring to the Case II for the time indicated in the figure; excitation laser line: $\lambda = 514.0$ nm.

The Raman spectra (excitation wavelength λ = 514.0 nm) of undegraded PVC1 and degraded at 60°C and 80°C at 60% RH are presented in **Figure 5** and the same temperatures at lower humidity are in **Figure S6.** The bands corresponding to polyene began to appear after a six-week degradation at 80°C, and after 12 weeks, the bands from polyenes were discernible at 60°C as well. The first peak at 1100 cm$^{-1}$ (denoted C-C=C) and 1514 cm$^{-1}$ (denoted C-C=C) originated from $v_1$(C-C) and $v_2$(C=C) stretching vibrations of polyenes [18]. Both C-C and C=C bonds exist in a sequential configuration known as conjugation.

As the quantity of HCl extracted from the backbone of the polymer chain increased, the PVC samples became more yellow, and the Raman intensity corresponding to polyene bands increased. Polyenes are strong Raman scatters when a 514 nm excitation laser line is used due to the strong electron/vibration coupling, and their vibrational modes can be selectively enhanced by resonance excitation nearby or close to the resonance with the $S_0 \rightarrow S_2$ (π–π*) electronic transition of polyene [57,63–65]. However, the appearance of polyene peaks in Raman spectra depends on several factors, including polyene content and excitation conditions. According to Kuznetsov et al. [66], the excitation wavelength λ (nm) corresponding to the optimal resonance Raman enhancement is a function of the polyene length *n*, as defined by **Equation 1**:

$$\lambda = 700 - 537.7\, e^{-0.0768n} \quad \textbf{(1)}$$

For the Raman spectrum recorded for degraded PVC1 under excitation at λ = 514.0 nm (green laser), polyenes with $n \approx 14$ exhibit the strongest resonance enhancement. According to **Equation 1**, a minimum of 32 polyene units must be present for resonance enhancement to occur as the wavelength shifts towards the red laser range. The currently employed amplification facilitates the detection of polyene; however, it concurrently obscures other signals, consequently rendering the detection of non-resonant species such as C-Cl bonds. For better detection of C-Cl bonds, it is generally more effective to utilize longer wavelength lasers, such as red one [31]. While Raman can detect the presence of polyenes and their vibrational modes, it is not quantitative in terms of counting the number of polyene structures.

Therefore, UV-Vis was employed in this study to determine the exact number of double bonds. Conjugated double bonds are clearly visible in UV-Vis spectrophotometry as they cause a redshift of the spectrum – plain PVC, with strictly saturated chains, has one absorption maximum at about 203 nm [65], while degraded PVC shows many different maxima, each of them corresponding to a specific number of conjugated double bonds [67]. Additionally, the use of a second derivative allows for better separation of adjacent maxima – they become minima with zeros of the function on either side of it.

Pristine PVC is colorless because its absorption maximum lies in the ultraviolet region of the electromagnetic spectrum [65,68]. With the creation of conjugated double bonds, the maximum undergoes a bathochromic shift (by the Woodward-Fieser and Fieser-Kuhn rules). When the number of conjugated double bonds is high enough (above 8 bonds [67] with sufficiently high statistical presence), the polymer starts absorbing light in the visible part of the spectrum, hence the color of the sample changes. An additional effect is the opaquing of the material (the rise in overall absorption, in the whole spectrum) due to different changes undergoing in the material, including reactions with the stabilizer, with oxygen from the air, etc. [68].

In the case of samples degraded at 60°C, RH = 60%, only small absorption in the visible light was detected, so the color change is nearly not detectable. Nevertheless, maxima from the ultraviolet part of the spectrum are shifting to the longer wavelength with the degradation time though, which indicates that the degradation progresses even at low temperatures. Assuming that the visible part of the electromagnetic spectrum begins at around 400 nm, the first maximum affecting the color of PVC1 is the one at 419 nm, corresponding to 9 conjugated double bonds [67] – it was present in all investigated samples. In PVC1 degraded at 60°C, RH = 60%, it is possible to notice maxima up to 12 conjugated double bonds with sufficient intensity to be clearly distinguishable from the noise (**Figure 6a, 6b**). Similar changes were observed for samples degraded at 60°C, RH = 20%, showing up to 12 conjugated double bonds present in the samples (**Figure S7a, S7b**).

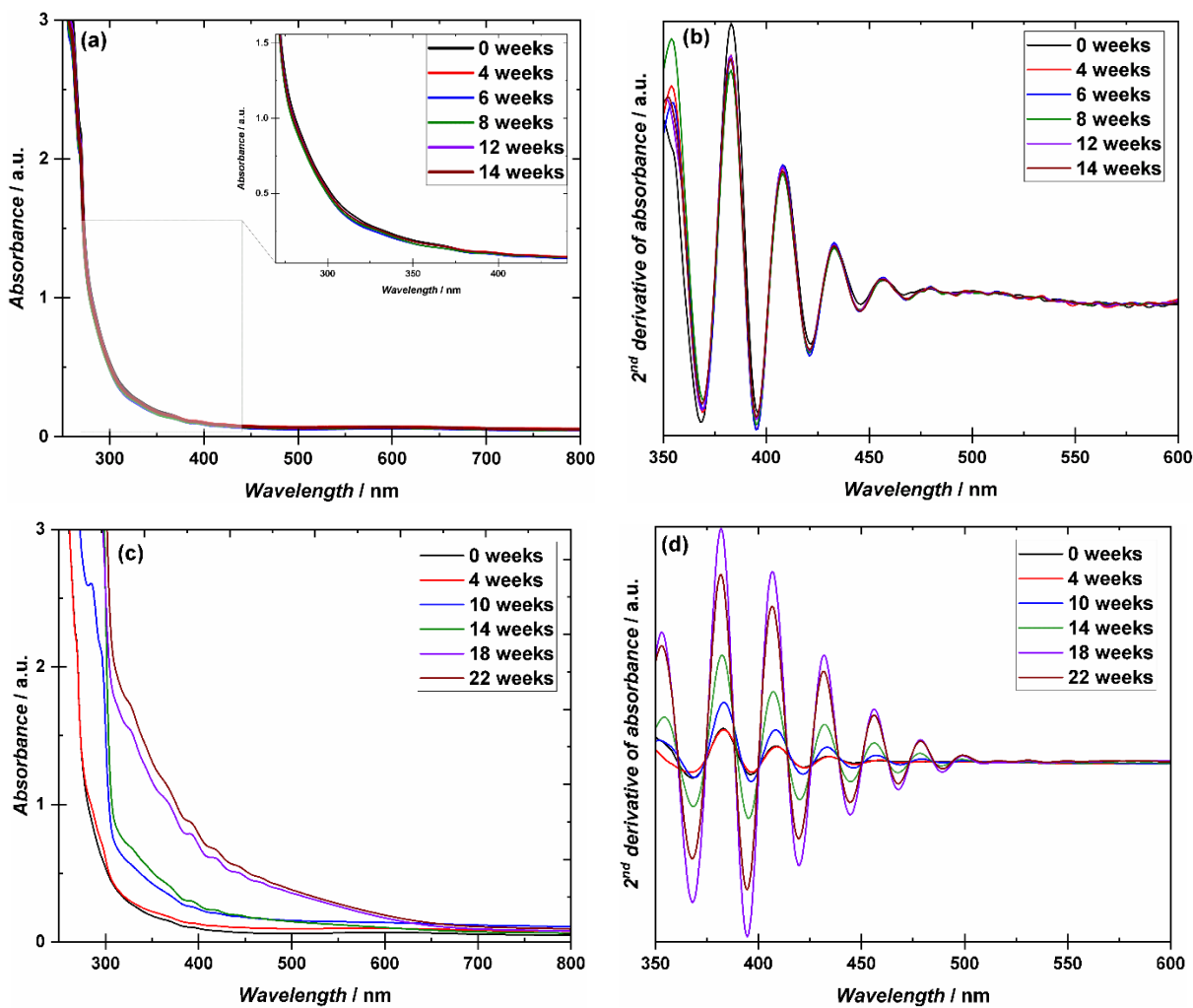

**Figure 6.** UV-Vis absorption spectra and the 2$^{nd}$ derivatives of smoothed spectra for PVC1 samples degraded at 60°C (**a** and **b**, respectively) and 80°C (**c** and **d**) with RH = 60% (Case II).

The spectra of samples degraded at 80°C, RH = 60%, correlate clearly with the color-change measurements – with the increase of degradation time, the number of conjugated double bonds increases simultaneously with the rise of the number of polymer chains having more double bonds. Only after 10 weeks of degradation does the absorption maxima start to have a high enough intensity in the visible part of the electromagnetic spectrum to cause a change of the sample's color (a visible yellowing) that then progresses with time. For the sample degraded for 20 weeks, the last absorption maximum on the spectrum with sufficient intensity to be observed, located at 560 nm, corresponds to 16 conjugated double bonds (**Figure 6c, 6d**). For the samples with an apparent change in their appearance, the absorbance at lower wavelengths exceeded the detector limit, resulting in a cut-off of some of the spectra for the first few absorption maxima. Samples degraded at 80°C without humidity control (**Figure S7c, S7d**) showed similar but slightly smaller changes, and the maximum number of conjugated double bonds was determined as 15 (last absorption maximum at 540 nm).

When the degradation temperature was raised to 120°C, the shift of absorption maxima progressed much more quickly than in lower temperatures, most probably due to the suspected mechanism change of degradation from ionic to radical. The most bathochromic shifted maximum, at around 516 nm, corresponds to 13 conjugated double bonds for the sample degraded for 5 days (**Figure S7e, S7f**).

Utilizing molar absorption coefficients of conjugated double bonds in degraded PVC described in the literature [69], as well as measured absorptivity values and the thickness of samples, the concentration of such moieties can be calculated using Beer-Lambert law. At 60°C (**Figure S8a, S8b**), the values are low and oscillate mostly due to sample differences and noise values. From absorptivity values at 80°C (**Figure S8c, S8d**), a clear trend of rising concentration of conjugated double bonds is drawn. Similarly, at 120°C (**Figure S8e**), the concentrations grow with degradation time, causing discoloration of samples. The comparison of values for samples degraded at the same temperature with different relative humidity suggests that the higher the humidity value, the faster the degradation progresses, leading to the creation of more conjugated double bonds.

Using the high linearity of absorption coefficients described by Baltacioğlu and Balköse [69], molar absorption coefficients for maxima corresponding to the conjugated polyene chain length of 11, 12, 13, and 14, as well as for two conjugated double bonds and a single double bond, were calculated by extrapolation (**Figure S9**). Utilizing both the extrapolated coefficients as well as the ones described in the literature, the total concentration of double bonds in the PVC matrix was calculated for the samples (**Table S3**). Additionally, with reference to the density and molar mass of the PVC repeating unit, a percentage of double bonds in the whole polymer was estimated. Moreover, the full dehydrochlorination of PVC results in about 58% mass loss of the sample (change in molar mass from 62.5 g/mol to 26 g/mol), so the percentage presence values of double bonds in the samples were further converted into approximate mass loss of the samples. The estimated values (**Figure S10**) are relatively low, indicating that even a small number of conjugated double bonds is sufficient to create visible changes. In Case II (RH = 60%) after 20 weeks at 60°C and 80°C, the resulting mass change is approximately 0.24% and 0.34%, respectively. This corresponds only to 0.0030% at 60°C, but 0.048% at 80°C of bonds being double bonds in the visible range. That makes the main difference in the investigated samples not the total concentration of double bonds, but the fraction of them making conjugated sequences absorbing in the visible part of the light, leading to color change. For samples degraded at lower humidity, Case I, the relative mass change reached 0.17% after 14 weeks of degradation at 60°C, RH = 20%, and 0.24% after 22 weeks of degradation at 80°C without humidity control, with 0.0022%/0.014% of bonds being double bonds in the visible spectrum for the most degraded samples, respectively. Additionally, non-zero numbers of double bonds were calculated for the undegraded sample too (0.18% mass change vs monomer and 0.0021% of bonds being double bonds in the visible spectrum), meaning the PVC foils undergo degradation even during the brief but harsh manufacturing and consecutive transportation and storage in variable conditions.

Assuming that the samples consist of mainly pure PVC polymer, with only a small percentage of additives or impurities, the remaining mass percentage, after subtracting the carbon, hydrogen, and nitrogen contents determined by elemental analysis (**Table S4**), can be equivalent to the chlorine content in samples. Then, the chlorine content change can be evaluated by comparing it to the undegraded sample – the calculated values are considerably low, no higher than 1%, which is similar to other results, including UV-Vis and Raman spectroscopies.

### 3.3. Surface changes in unplasticized PVC

Contact angle measurements investigated the changes in surface polarity/hydrophobicity during thermal degradation. The water contact angle measures non-covalent forces between water and the first monolayer of PVC. Thus, in the case of strong attractions (adhesion) between phases (hydrophilicity of the surface), the water drops spread on the PVC and wet it, rendering lower values of water contact angle (WCA). The surface of pristine PVC is mostly rich in C–C and C–H groups and is thus hydrophobic. The hydrophobicity of PVC remains unchanged for Case II when PVC1 was degraded at 60% RH, both at 60°C and 80°C, as shown in **Figure 7**. However, the decrease in hydrophobicity presents clearly in Case I of PVC degradation at 80°C, with no humidity control as shown in **Figure S11**, which can be attributed to the prevalence of the ionic mechanism, which is also the reason for the slow degradation.

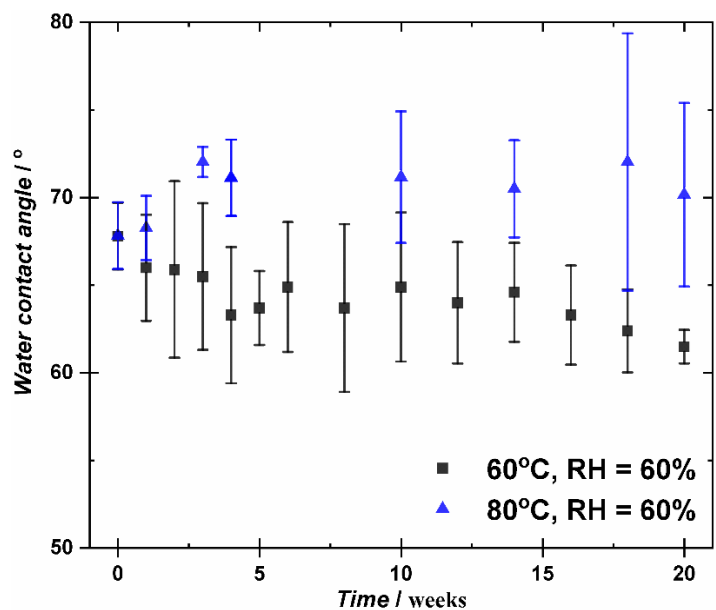

**Figure 7.** The contact angle of water over PVC1 degraded at 60°C and 80°C with RH = 60%.

The XPS spectra of Cl 2p for both undegraded and degraded PVC at 60°C and 80°C with RH equal to 60% for 20 weeks (Case II) are provided in **Figure 8**. The Cl 2p spectra for PVC have three peaks: Cl $2p^{1/2}$ at 198.49 eV peak assigned to chloride ion and Cl $2p^{3/2}$ at 200.58 eV and 202.20 eV that represent organic chlorine atoms covalently bounded sp2 carbon (C–Cl).

The Cl 2p signal remains unaltered for the PVC sample subjected to degradation at 60°C and 60% relative humidity. However, as can be seen in **Figure 8c**, there is a considerable decrease in the Cl 2p signal for the PVC1 sample degraded at 80°C and 60% relative humidity in comparison to the undegraded sample. The reduction in Cl 2p intensity, along with an increase in the amount of carbon as detected by elemental analysis (**Table S4**), is attributed to the formation of polyenes, as confirmed by Raman and UV-Vis spectroscopy. Additionally, under these conditions, the radical mechanism of polyenes formation might be at work, hence the presence of radicals was detected by EPR. Furthermore, under these conditions, the radical mechanism of polyene formation may be operative, as evidenced by the detection of radicals by EPR.

In Case I, referring to low RH degradation conditions, after 20 weeks with no humidity control (**Figure S12**), the Cl 2p peak splits into two components only, one at 200.69 eV, assigned to covalent chlorine, and one peak at 199.09 eV, attributed to the formation of more ionic chlorine which can clarify the increase in the hydrophilicity of the PVC surface.

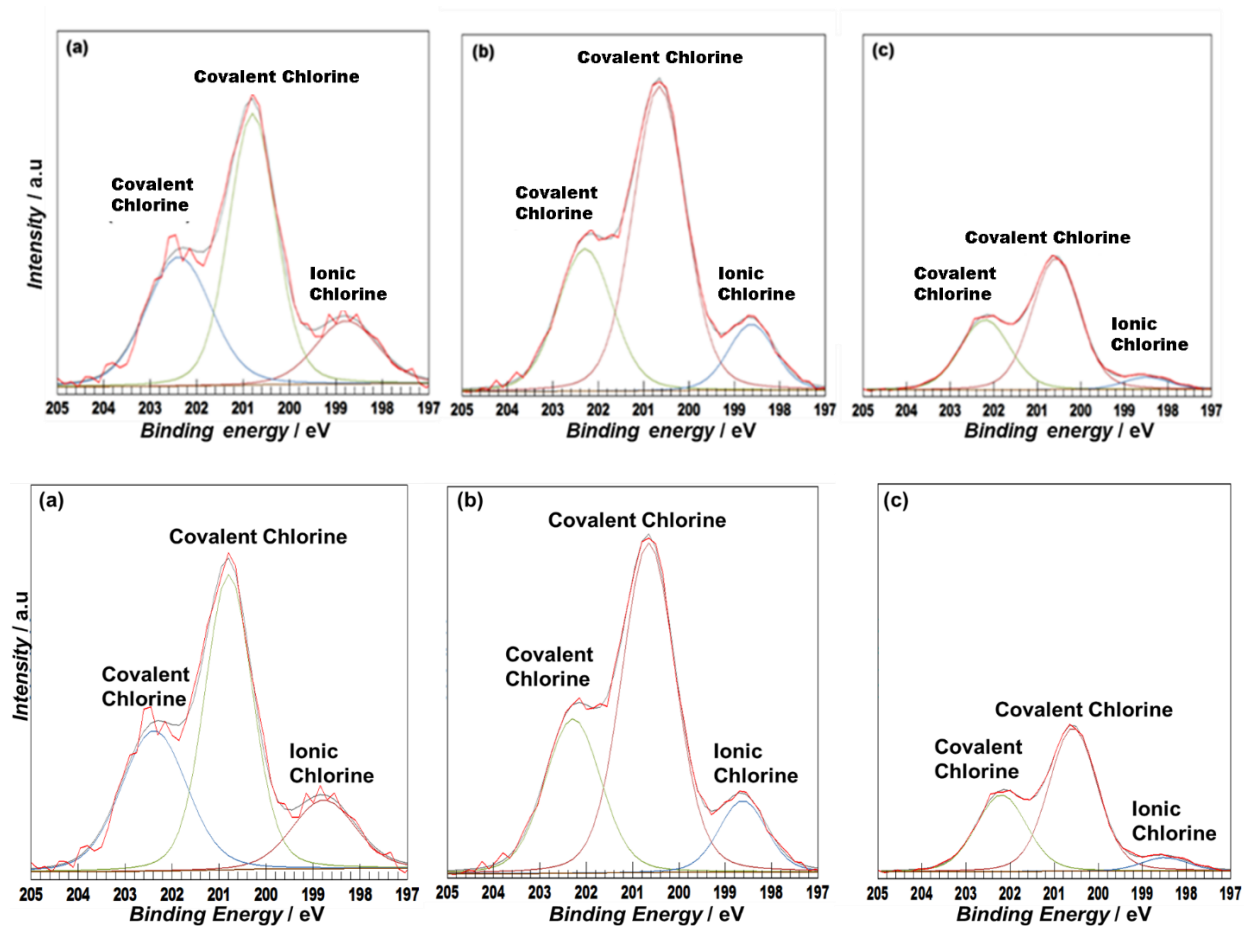

**Figure 8.** The XPS Cl 2p$^{3/2}$ and Cl 2p$^{1/2}$ spectra of undegraded PVC1 **(a),** PVC degraded at 60˚C **(b)** PVC degraded at 80˚C **(c)** for 20 weeks at RH = 60% (Case II).

Notwithstanding the fact that some chlorine leaves the PVC chain, thereby causing polyene formation and sample yellowing, it can be stored in the sample as ionic chlorine, which is trapped by the polyene, as proposed in **Scheme 2**. The decrease in both covalent and ionic chlorine at 60% relative humidity (RH) and 80°C indicates that humidity influences the overall degradation behavior of PVC. This effect probably facilitates the removal of hydrogen chloride from the polymer matrix, as presented earlier in **Scheme 1.** Furthermore, the appearance of a radical signal in EPR spectra after degradation suggests radical (nonionic) pathways of degradation.

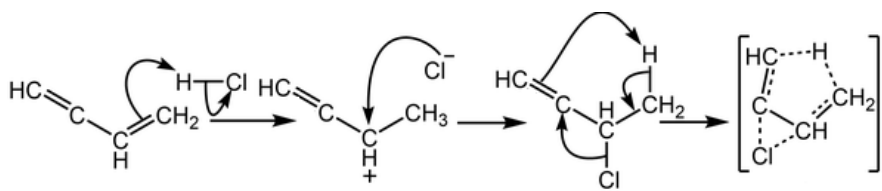

**Scheme 2.** Trapping of chlorine ion by polyene formed in PVC chain.

Evaluation of the hydrogen chloride emission with the Oddy test showcased that, even though the amount of chlorine removed from the PVC chains was low, when released it was sufficient to cause changes in the testing metal foils. For silver (**Figure S13a**) and lead (**Figure S13c**) coupons, no changes were observed, but for copper (**Figure S13b**) foils, the changes were possible to be observed with the naked eye. The surface of copper foils was covered in green spots signifying the formation of copper chloride, and a red tarnish to some degree was also noted. The changes point to the conclusion that the investigated PVC1 foil should receive the "temporary" rating in the Oddy test for the copper coupons and the "permanent" rating for the other foils, suggesting that rigid PVC is safe for storage near but not in contact with other art pieces for up to six months due to the possibility of degradation caused by released HCl.

## 3.4. Evolution of mechanical properties

To find the link between the chemical degradation of the PVC leading to discoloration and changes in mechanical properties, non-destructive dynamic mechanical analysis (DMA) was used to study changes in the materials' properties of the PVC1 samples with degradation time. Generally, DMA experiments provide information on the elastic and viscous response of the material expressed as the storage $E'$ and loss $E''$ moduli, respectively. Storage modulus, representing the elastically stored energy under external load, corresponds to the stiffness of the sample, i.e. the higher the $E'$, the stiffer the material. Due to variability in the obtained results highlighted by the error bars, experimental error determined for each point was treated as the weight for the linear fitting procedure and ANOVA analysis was applied to evaluate the statistical significance of the resulting slope.

In Case I of degradation (low relative humidity up to 30%, **Figure S14**), the slopes of the linear functions were not statistically different from 0 (at the significance level of $\alpha = 0.05$). Thus, the stiffness of the PVC1 remained unaffected by the structural changes occurring due to the dehydrochlorination reaction over degradation time. **Figure 9** shows the time evolution of the $E'$ determined in the oscillatory test at a frequency of 1 Hz for PVC1 samples degraded at 60% RH (Case II). Similarly to Case I, no evidence of PVC1 stiffening was observed when degraded at 60°C, RH = 60% (**Figure 9a**). On the contrary, when degradation was carried out at 80°C, RH = 60% (**Figure 9b**), stiffening of the PVC1 was observed as indicated by the statistically significant slope ($p = 0.003$) and resulting relation of $E'=f(t)$:

$$E' = 0.006t + 1.315 \qquad (2)$$

where $E'$ is storage modulus (stiffness) in GPa and $t$ is degradation time in days.

Using previously determined activation energy value for dehydrochlorination at mild degradation conditions, i.e. 86 kJ mol$^{-1}$ [32], prediction of $E'=f(t)$ for 20°C and 60% RH is:

$$E' = 1.5*10^{-5}t + 1.315 \qquad (3)$$

which indicates that expected stiffening is 400-fold slower at typical display conditions in comparison to accelerated degradation performed in this study (at 80°C, RH = 60%).

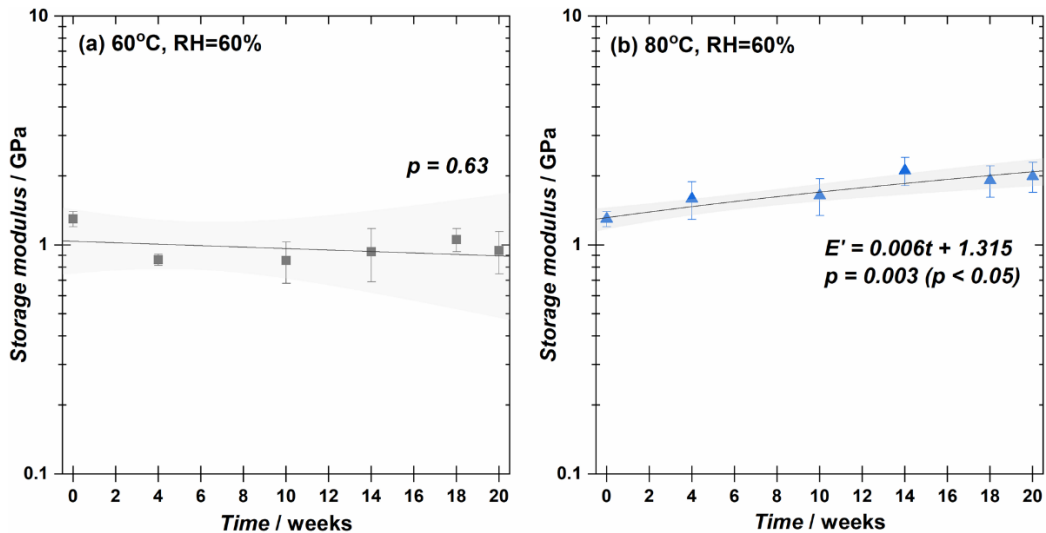

**Figure 9.** Storage modulus $E'$ evolution with degradation time for the PVC1 degraded at 60°C, RH = 60% **(a)** and 80°C, RH = 60% **(b)** corresponding to Case II. The grey area represents confidence bands corresponding to the confidence level of 95%.

Generally, stiff objects are more brittle and thus more prone to mechanical damage caused by the high impact, e.g. during transportation. To assess the potential risk of mechanical damage related to discoloration, the damage function describing the yellowing rate proposed by Rijavec et al. was applied [32]. Assuming constant average conditions of 20°C and 60% RH, the expected time to just noticeable yellowness change determined by $\Delta b = 1.5$ [32] for unplasticized PVC ($M_w$ of ca. $10^5$ g mol$^{-1}$) is ca. 3.5 years and the accompanying increase in stiffening will be of ca. 1.5%. In the case of yellowing defined by $\Delta b = 5$ considered as a different color [39], the lifetime of the unplasticized PVC is ca. 12 years and the expected increase in the material's stiffness will be of ca. 5.0% in this time frame. Correlation between the rate of yellowing and rate of stiffening for unplasticized PVC matrix revealed that even extensive discoloration of the PVC leads to a small change in the macroscopic mechanical properties of the material, thus the thermal degradation and resulting yellowing of the PVC artifacts does not contribute significantly to a higher risk of mechanical damage due to large external loads (impact) during handling or transportation.

The results for Case I as well as Case II are in good agreement with molecular weight distribution analysis. Thermal degradation at 80°C without humidity (Case I) control for up to 18 weeks did not induce statistically significant changes in chain length or macromolecular architecture (e.g. crosslinking) as can be seen in **Figure S15**. Also, in the Case II (degradation at 80°C and RH = 60%), SEC results do not indicate significant chain scissions or crosslinking for samples aged for 18 weeks (**Figure S15**). This is in agreement with our previous study [30], which also found no statistically significant changes in the molecular weights of PVC1 samples after thermal degradation (50°C and 70°C) at elevated humidity levels (RH = 80%). The findings presented here are generally in agreement with other literature reports, which demonstrate that artificial thermal degradation at 80°C (performed for up to 35 weeks) [70] and at 100°C for 4 days [71] does not affect the length of PVC macromolecular chains.

## 4. Conclusion

The thermal degradation of unplasticized PVC at moderate temperatures and under different relative humidities has been extensively studied. The findings indicated that not all analytical methods were sensitive to the changes in PVC structure that occurred as a result of degradation. The most conspicuous consequence of the thermal degradation of PVC is discoloration, mostly observed as a yellowing of the material. The formation of polyene chains within the PVC structure is responsible for the observed color change, which is confirmed by Raman and UV-Vis spectroscopies. In the context of Raman spectroscopy, it was demonstrated that the selection of an appropriate excitation wavelength is of paramount importance. Furthermore, radical degradation mechanisms must be considered, even at relatively low temperatures, as they play a pivotal role in PVC degradation under conditions of high humidity (Case II). Conversely, only ionic mechanisms are active at low humidity (Case I), exerting additional influence on the degradation process. Despite the extensive discoloration, the macroscopic mechanical properties of PVC are only marginally affected. These findings underscore the necessity for cautious environmental management with low temperatures and low humidity to ensure the preservation of both the appearance and structural integrity of PVC objects. Furthermore, this paper provides an overview of the various experimental methods that facilitate the recommendation of the most suitable techniques, offering valuable insight throughout the investigation of PVC degradation. These include colorimetry, Raman, UV-Vis, EPR, and XPS techniques. In the context of heritage and conservation science, the majority of these techniques are non-destructive, rendering them preferable for on-site studies and collection surveys in museums and galleries.

**Acknowledgments**

The authors would like to express their gratitude to Dr. Monika Gołda–Cępa and Ms. Paulina Chytrosz-Wróbel from Jagiellonian University for their assistance with contact angle measurements, MS. Jernej Imperl and MS. Daša Terobšič from University of Ljubljana for their assistance with metal content determination, and Prof. Ing. Stepan Podzimek for performing SEC analysis of PVC1 sample degraded at 80ºC and RH = 60%.

The research was carried out within the OPUS LAP 20, National Science Centre, Poland, grant no. 2020/39/I/HS2/00911 and Slovenian Research and Innovation Agency, N1-0241 project, CEUS scheme, as well as projects P1-0447, J7-50226 and I0-E012. The infrastructure that was funded by the European Union in the framework of the Smart Growth Operational Programme, Measure 4.2, Grant No. POIR.04.02.00-00-D001/20, was used.

**Supplementary Materials for:**

**Effect of Accelerated Thermal Degradation of Poly(Vinyl Chloride): The Case of Unplasticized PVC**

Marwa Saad[1], Marek Bucki[2], Sonia Bujok[3], Dominika Pawcenis[1], Tjaša Rijavec[4], Karol Górecki[1], Łukasz Bratasz[3], Irena Kralj Cigić[4], Matija Strlič[4], Krzysztof Kruczała[1*]

[1] Faculty of Chemistry, Jagiellonian University in Krakow, Gronostajowa 2, 30-387 Krakow, Poland

[2] Jagiellonian University, Doctoral School of Exact and Natural Sciences, Gronostajowa 2, 30-387, Krakow, Poland

[3] Jerzy Haber Institute of Catalysis and Surface Chemistry, Polish Academy of Sciences, Niezapominajek 8, 30–239 Kraków, Poland

[4] Faculty of Chemistry and Chemical Technology, University of Ljubljana, Večna pot 113, 1000 Ljubljana, Slovenia

E-mail: kruczala@chemia.uj.edu.pl

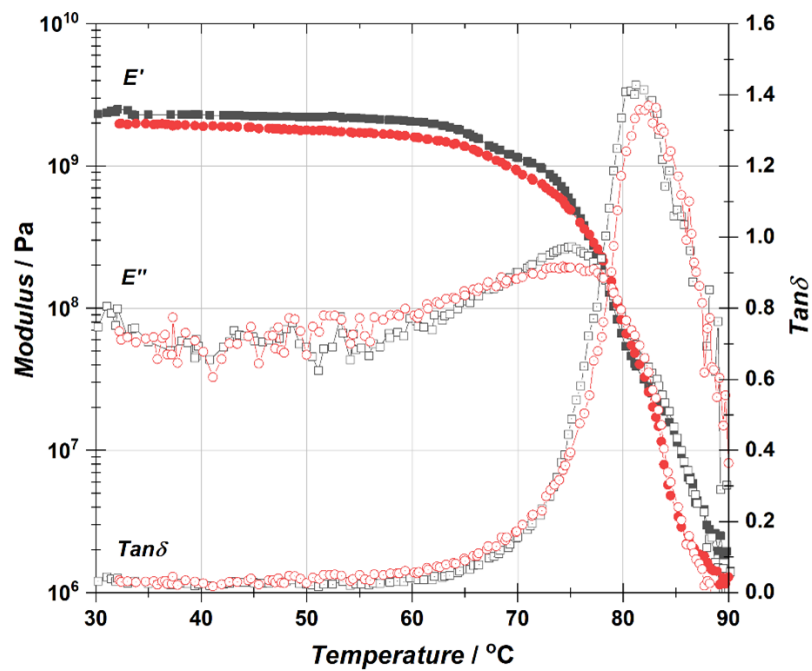

**Figure S1.** Temperature dependence of storage ($E'$) and loss ($E''$) modulus together with tan$\delta$ of the undegraded (pristine) PVC1 (measured twice – black and red datasets).

Dynamic mechanical thermal analysis (DMTA) enables the investigation of the glass transition and determination of $T_g$ of polymer materials as a clear change in relation of storage and loss moduli given by the tan$\delta = E''/E'$. Herein, the maximum of the tan$\delta$ peak is considered as the $T_g$ determinant. Thus, below $T_g$ (81.8°C), PVC1 behaves like a glassy viscoelastic solid, whereas above $T_g$ it becomes a rubbery solid. Worth mentioning is that glass transition occurs in a range of temperatures where the segmental motion of macromolecules increases, thus in the case of PVC1 it starts around 60°C and ends presumably around 100°C, however the range of glass transition is strongly affected by the dispersity of the polymer, method used for $T_g$ determination, and most importantly the rate of heating (e.g. DSC curves obtained typically at 10°C min$^{-1}$ heating rate would indicate narrower glass transition range than DMTA curves obtained at 2°C min$^{-1}$, resulting in a wider glass transition temperature range which is observed in **Figure S1**).

**Table S1.** The elemental composition of pristine PVC1 sample, measured by ICP-OES and ICP-MS after microwave-assisted mineralization.

| Method | ICP-OES | | ICP-MS | |
|---|---|---|---|---|
| Element investigated | Concentration /ppm | SD /ppm | Concentration /ppm | SD /ppm |
| Sn | 1594 | 43 | 1022 | 70 |
| Na | 88 | 39 | 105.4 | 1.1 |
| Ca | 73 | 17 | 71.7 | 3.9 |
| Al | 24 | 2 | 65.9 | 1.1 |
| K | 40 | 22 | 12.1 | 0.5 |
| Fe | 10 | 3 | 9.01 | 0.17 |
| Mg | - | - | 1.83 | 0.03 |
| Zn | - | - | 0.29 | 0.01 |

**Table S2.** The evolution of $\Delta L$, $\Delta a$, $\Delta b$, and $\Delta E$ as a function of the degradation time for PVC1 at low RH - Case I and RH=60% - Case II.

| Time / week | $\Delta L$ | $\Delta a$ | $\Delta b$ | $\Delta E$ |
|---|---|---|---|---|
| *Case I: Low relative humidity (up to RH = 30%)* | | | | |
| 60°C, RH = 20% | | | | |
| 2 | -2.46 | 0.04 | 0.11 | 1.62 |
| 4 | -2.52 | 0.06 | 0.37 | 1.69 |
| 8 | -2.43 | 0.02 | 0.32 | 1.62 |
| 12 | -2.27 | 0.03 | 0.27 | 1.51 |
| 14 | -2.27 | 0.09 | 0.12 | 1.50 |
| 70°C, RH = 30% | | | | |
| 1 | -0.75 | 0.03 | 0.37 | 0.57 |
| 2 | -1.02 | 0.19 | 0.71 | 0.94 |
| 3 | -1.02 | 0.1 | 0.58 | 0.83 |
| 4 | -1.28 | 0.16 | 0.90 | 1.16 |
| 5 | -1.34 | 0.10 | 0.95 | 1.20 |
| 6 | -1.56 | 0.09 | 0.90 | 1.52 |
| 7 | -1.61 | 0.1 | 1.65 | 1.78 |
| 8 | -1.97 | 0.29 | 2.50 | 2.55 |
| 9 | -2.18 | 0.06 | 2.70 | 2.71 |
| 80°C, No humidity control (RH < 7%) | | | | |
| 2 | -0.73 | 0.13 | 0.30 | 0.59 |
| 4 | -0.95 | 0.09 | 0.87 | 1.02 |
| 6 | -1.83 | 0.03 | 1.54 | 1.83 |
| 8 | -2.49 | 0.14 | 3.32 | 3.29 |
| 10 | -3.03 | 0.31 | 5.75 | 5.11 |
| 12 | -3.60 | 0.23 | 9.33 | 7.56 |
| 14 | -5.63 | 0.89 | 13.95 | 10.82 |
| 16 | -8.30 | 2.46 | 16.76 | 13.29 |
| 18 | -17.65 | 11.65 | 33.63 | 24.07 |
| 20 | -19.16 | 13.64 | 36.06 | 25.52 |
| *Case II: Typical top limit of RH in heritage institutions – 60% RH* | | | | |
| **60˚C, RH = 60%** | | | | |
| 2 | -1.11 | 0.08 | 0.06 | 0.81 |
| 4 | -1.15 | 0.08 | 0.23 | 0.86 |
| 6 | -0.12 | 0.06 | 0.83 | 0.79 |
| 8 | -0.33 | 0.05 | 0.82 | 0.81 |
| 10 | -0.15 | 0.46 | 0.89 | 1.03 |
| 12 | -0.59 | 0.18 | 0.95 | 1.01 |
| 14 | -1.07 | 0.21 | 1.14 | 1.34 |
| 16 | -0.44 | 0.28 | 1.47 | 1.44 |
| 18 | -0.25 | 0.36 | 1.68 | 1.62 |
| 20 | -0.82 | 0.45 | 1.90 | 1.92 |
| **80˚C, RH = 60%** | | | | |
| 1 | -0.01 | 0.62 | 4.95 | 4.10 |
| 4 | -4.17 | 0.03 | 11.21 | 8.84 |
| 10 | -34.67 | 22.72 | 35.40 | 36.41 |
| 14 | -46.81 | 30.58 | 38.73 | 48.08 |
| 18 | -52.75 | 33.21 | 38.37 | 54.54 |
| 20 | -54.07 | 34.09 | 38.62 | 56.12 |
| 28 | -58.50 | 35.75 | 38.91 | 61.50 |
| 32 | -58.36 | 36.13 | 38.64 | 61.33 |

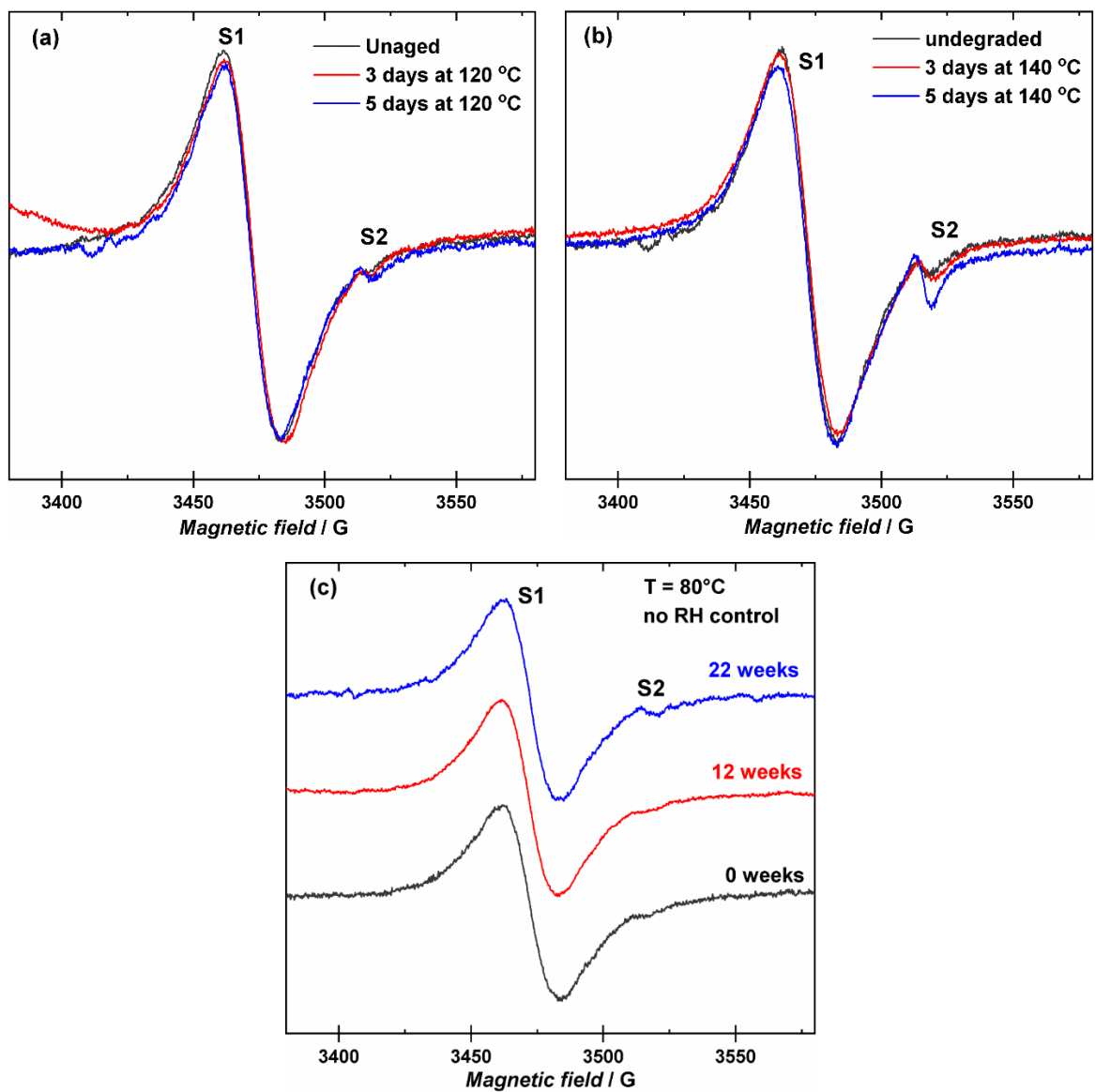

**Figure S2.** X-band EPR spectra for PVC1 degraded for 3 and 5 days at 120°C **(a)**, 140°C **(b),** and PVC1 degraded for 22 weeks at 80°C with no humidity control (**c** – example of Case I).

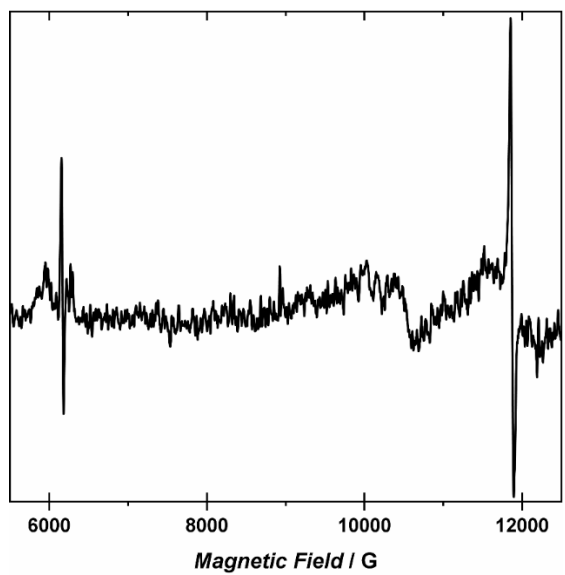

**Figure S3.** Q-band EPR spectrum of undegraded PVC1.

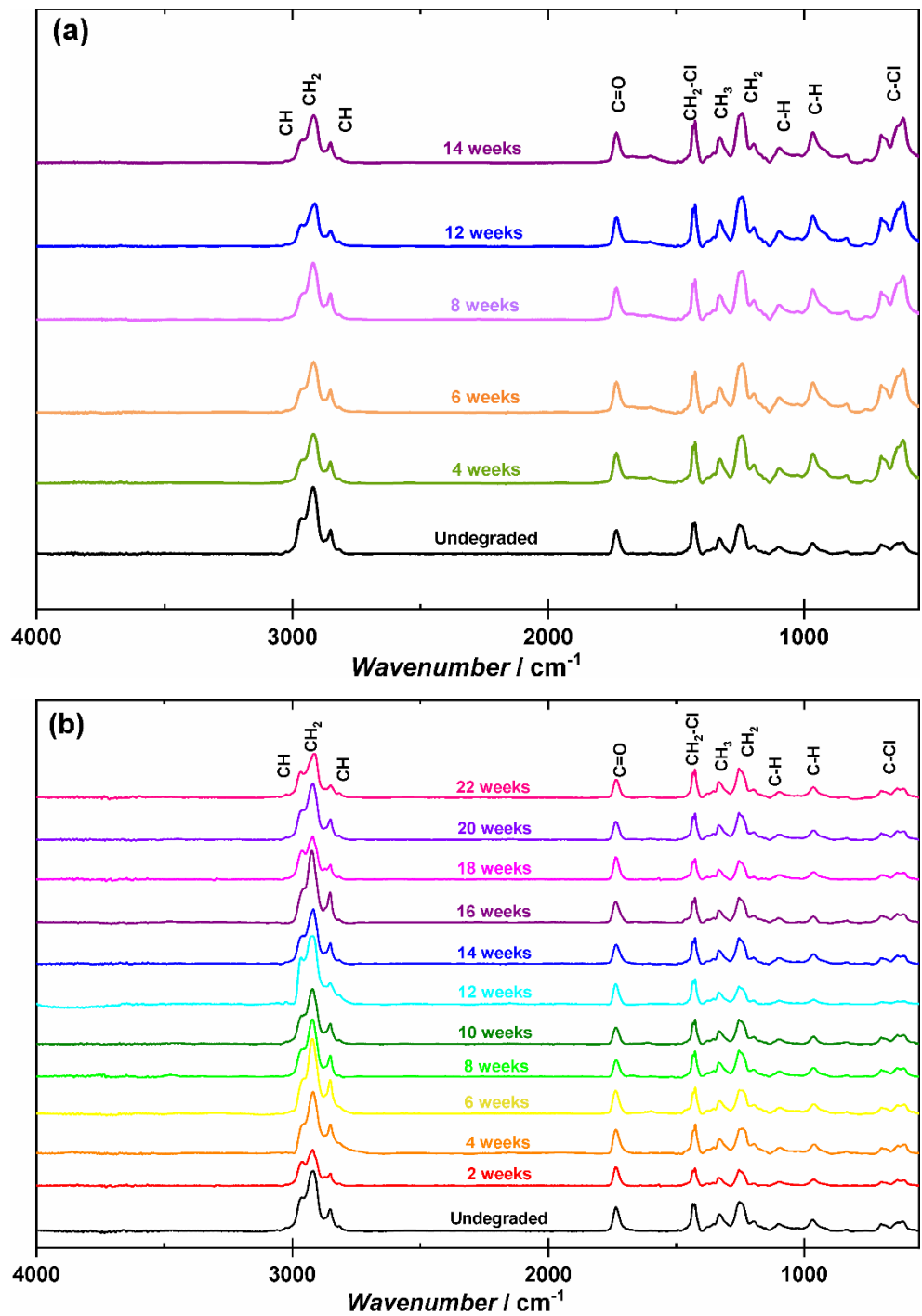

**Figure S4.** ATR-FTIR for undegraded and degraded PVC1 at 60°C, RH = 20% (**a**), 80°C in a dry chamber (no humidity control) (**b**) for the time indicated in the figure.

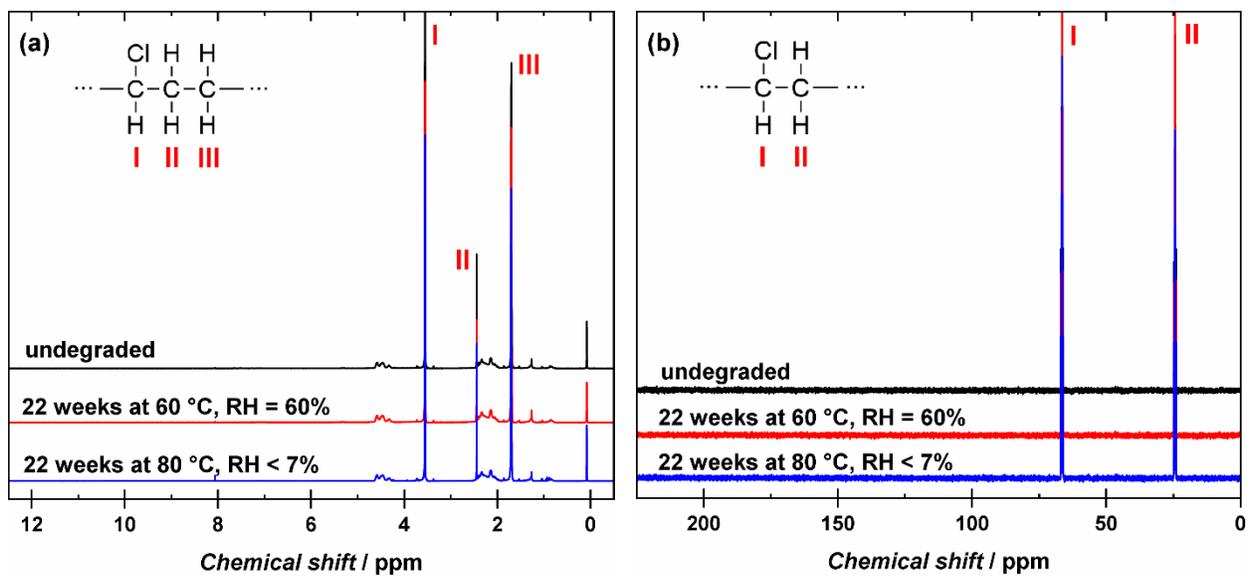

**Figure S5.** Comparison of $^1$H (**a**) and $^{13}$C (**b**) NMR spectra of PVC1 – undegraded (black), degraded for 22 weeks at 60°C, RH = 60% (red), and at 80°C, no RH control (blue).

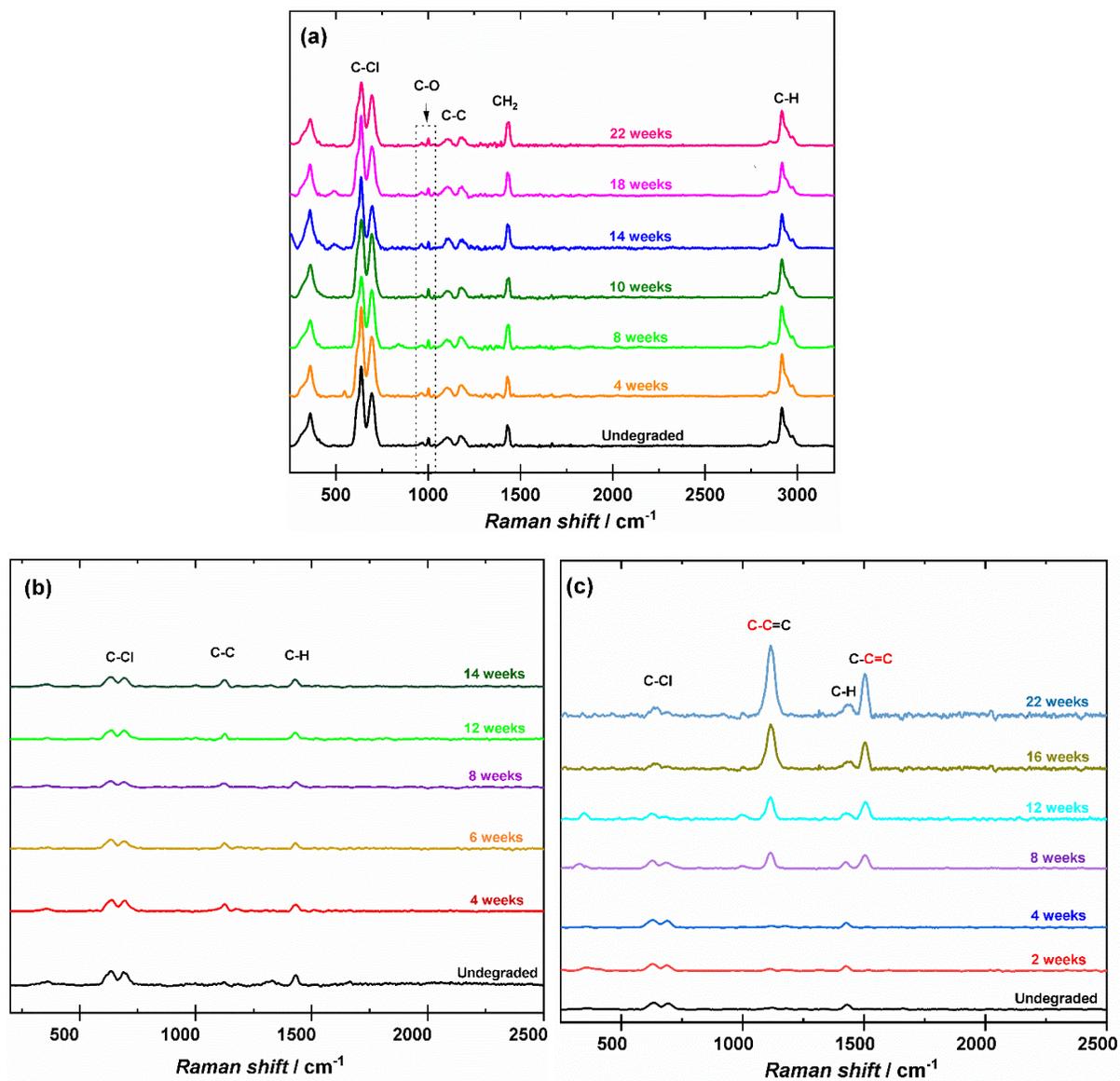

**Figure S6.** Raman spectra of PVC1 undegraded and degraded at 80°C in a dry chamber (no humidity control) for time indicated in the figure; excitation laser line λ = 785 nm (**a**); PVC1 undegraded and degraded at 60°C, RH = 20% for time indicated in the figure; excitation laser line λ = 514 nm (**b**); 80°C in a dry chamber (no humidity) for the time indicated in the figure; excitation laser line λ = 514 nm (**c**).

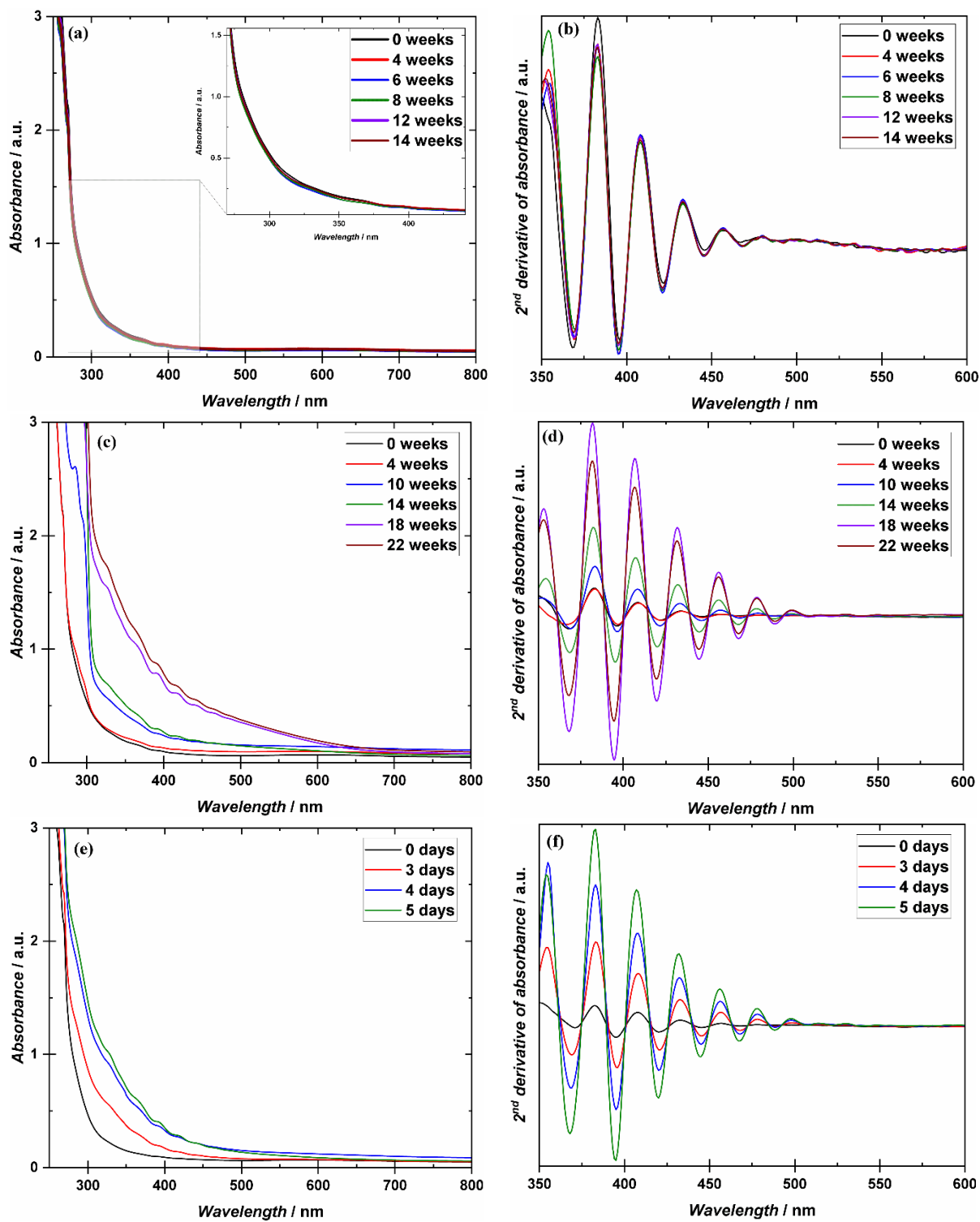

**Figure S7**. UV-Vis absorption spectra and the 2$^{nd}$ derivatives of smoothed spectra for PVC1 samples degraded at 60°C with RH = 20% (**a** and **b**) and 80°C without RH control (**c** and **d**) (Case I), as well as at 120 °C without humidity control (**e** and **f**).

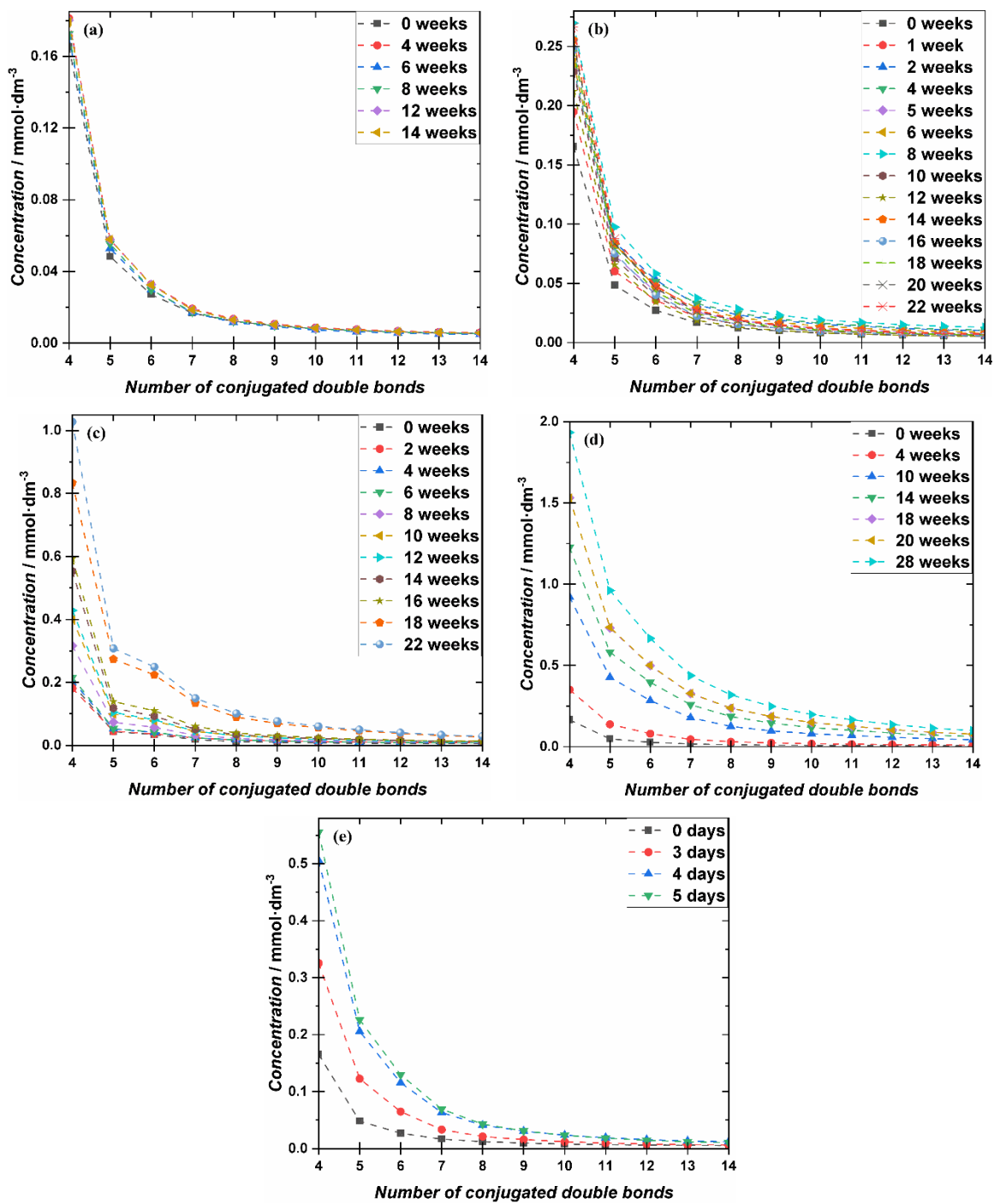

**Figure S8.** The concentration of conjugated double bonds for samples degraded at 60°C, RH = 20% (**a**) and 60% (**b**), at 80°C without humidity control (**c**) and with RH = 60% (**d**), as well as at 120°C (**e**), calculated using Beer-Lambert law.

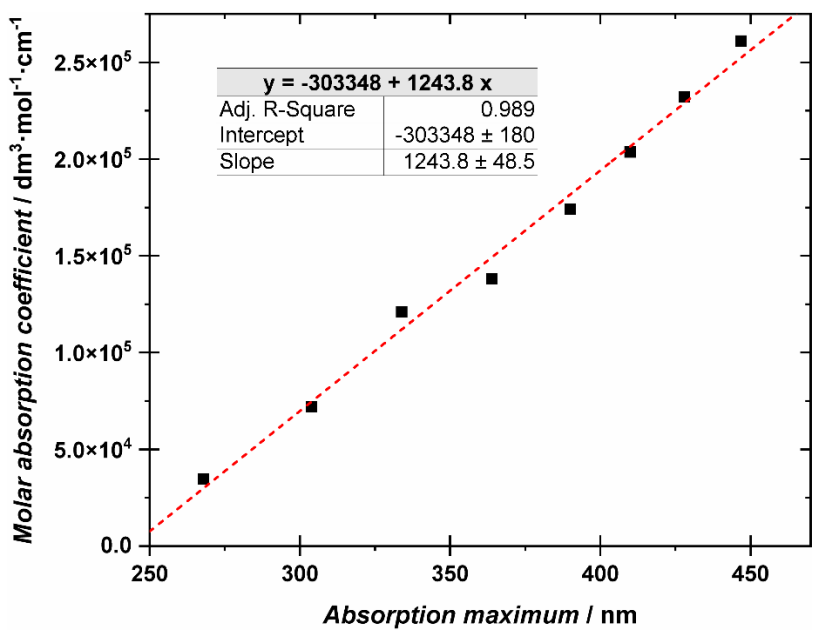

**Figure S9**. The linear regression of the relationship between molar absorption coefficients and the location of the absorption maximum on the spectrum.

**Table S3.** The changes in the concentration of double bonds, their percentage presence in the degraded samples, as well as the relative mass changes of the samples.

| *Case I: Low relative humidity (up to RH = 30%)* | | | | |
|---|---|---|---|---|
| **PVC1 degraded at 60°C, RH = 20%** | | | | |
| Degradation time /weeks | Total molar concentration of double bonds /mol·dm$^{-3}$ | Percentage of double bonds in the sample | Percentage of double bonds absorbing in the visible spectrum | Relative mass change due to dehydrochlorination |
| 0 | 0.069 | 0.31% | 0.0021% | -0.18% |
| 4 | 0.066 | 0.30% | 0.0023% | -0.17% |
| 6 | 0.063 | 0.29% | 0.0019% | -0.17% |
| 8 | 0.064 | 0.29% | 0.0021% | -0.17% |
| 12 | 0.066 | 0.30% | 0.0022% | -0.17% |
| 14 | 0.065 | 0.30% | 0.0022% | -0.17% |
| **PVC1 degraded at 80°C, no RH control** | | | | |
| Degradation time /weeks | Total molar concentration of double bonds /mol·dm$^{-3}$ | Percentage of double bonds in the sample | Percentage of double bonds absorbing in the visible spectrum | Relative mass change due to dehydrochlorination |
| 0 | 0.069 | 0.31% | 0.0021% | -0.18% |
| 2 | 0.064 | 0.29% | 0.0032% | -0.17% |

| Degradation time /weeks | Total molar concentration of double bonds /mol·dm$^{-3}$ | Percentage of double bonds in the sample | Percentage of double bonds absorbing in the visible spectrum | Relative mass change due to dehydrochlorination |
|---|---|---|---|---|
| 4 | 0.072 | 0.32% | 0.0032% | -0.19% |
| 6 | 0.077 | 0.35% | 0.0030% | -0.20% |
| 8 | 0.075 | 0.34% | 0.0038% | -0.20% |
| 10 | 0.070 | 0.32% | 0.0054% | -0.18% |
| 12 | 0.080 | 0.36% | 0.0047% | -0.21% |
| 14 | 0.084 | 0.38% | 0.0054% | -0.22% |
| 16 | 0.081 | 0.37% | 0.0059% | -0.21% |
| 18 | 0.092 | 0.41% | 0.013% | -0.24% |
| 22 | 0.092 | 0.41% | 0.014% | -0.24% |
| **PVC1 degraded at 120˚C, no RH control** | | | | |
| Degradation time /weeks | Total molar concentration of double bonds /mol·dm$^{-3}$ | Percentage of double bonds in the sample | Percentage of double bonds absorbing in the visible spectrum | Relative mass change due to dehydrochlorination |
| 0 | 0.069 | 0.31% | 0.0021% | -0.18% |
| 3 | 0.073 | 0.33% | 0.0030% | -0.19% |
| 4 | 0.084 | 0.38% | 0.0057% | -0.22% |
| 5 | 0.086 | 0.39% | 0.0054% | -0.23% |
| *Case II: Typical top limit of RH in heritage institutions – 60% RH* | | | | |
| **PVC1 degraded at 60˚C, RH = 60%** | | | | |
| Degradation time /weeks | Total molar concentration of double bonds /mol·dm$^{-3}$ | Percentage of double bonds in the sample | Percentage of double bonds absorbing in the visible spectrum | Relative mass change due to dehydrochlorination |
| 0 | 0.069 | 0.31% | 0.0021% | -0.18% |
| 2 | 0.065 | 0.29% | 0.0041% | -0.17% |
| 4 | 0.060 | 0.27% | 0.0043% | -0.16% |
| 5 | 0.066 | 0.30% | 0.0031% | -0.17% |
| 6 | 0.067 | 0.30% | 0.0036% | -0.18% |
| 8 | 0.071 | 0.32% | 0.0051% | -0.19% |
| 10 | 0.089 | 0.40% | 0.0026% | -0.23% |
| 12 | 0.086 | 0.39% | 0.0022% | -0.23% |
| 14 | 0.082 | 0.37% | 0.0032% | -0.22% |
| 16 | 0.092 | 0.42% | 0.0025% | -0.24% |
| 18 | 0.086 | 0.39% | 0.0027% | -0.23% |
| 20 | 0.086 | 0.39% | 0.0030% | -0.23% |
| 22 | 0.089 | 0.40% | 0.0030% | -0.24% |
| **PVC1 degraded at 80˚C, RH = 60%** | | | | |
| Degradation time /weeks | Total molar concentration of | Percentage of double | Percentage of double bonds | Relative mass change due to dehydrochlorination |

|   | double bonds /mol·dm$^{-3}$ | bonds in the sample | absorbing in the visible spectrum |   |
| --- | --- | --- | --- | --- |
| 0  | 0.069 | 0.31% | 0.0021% | -0.18% |
| 4  | 0.073 | 0.33% | 0.0046% | -0.19% |
| 10 | 0.106 | 0.48% | 0.0198% | -0.28% |
| 14 | 0.116 | 0.52% | 0.0290% | -0.31% |
| 18 | 0.121 | 0.55% | 0.0362% | -0.32% |
| 20 | 0.124 | 0.56% | 0.0366% | -0.33% |
| 28 | 0.130 | 0.59% | 0.0478% | -0.34% |

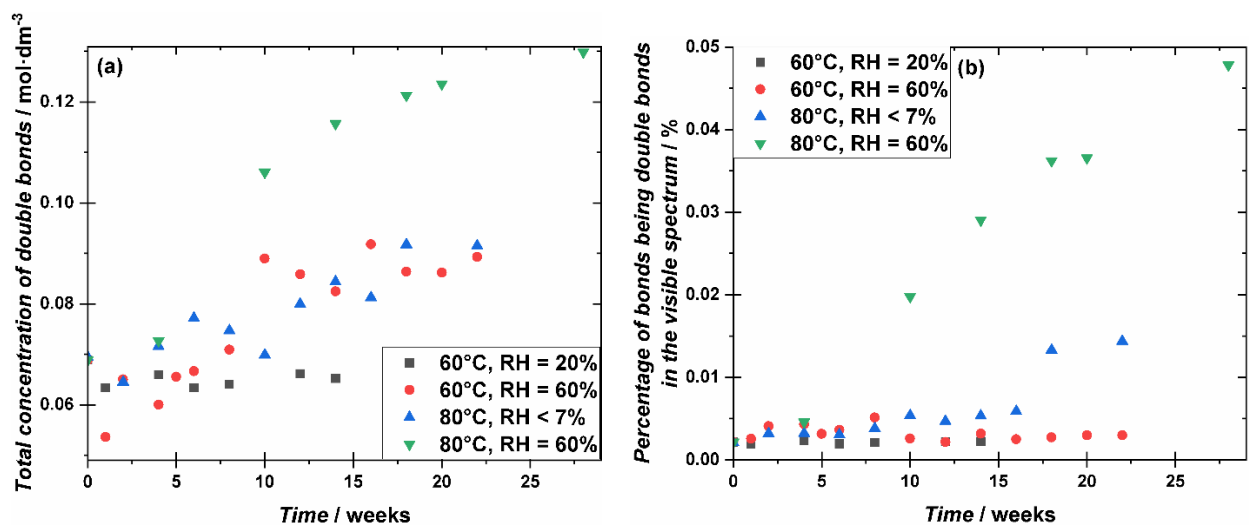

**Figure S10.** The total concentration of conjugated double bonds (**a**) and the percentage of all bonds being double bonds in the visible spectrum (**b**) for samples aged at 60°C, RH = 20% (**black**) or 60% (**red**) and 80°C, no humidity control (**blue**) or RH = 60% (**green**) plotted versus degradation time.

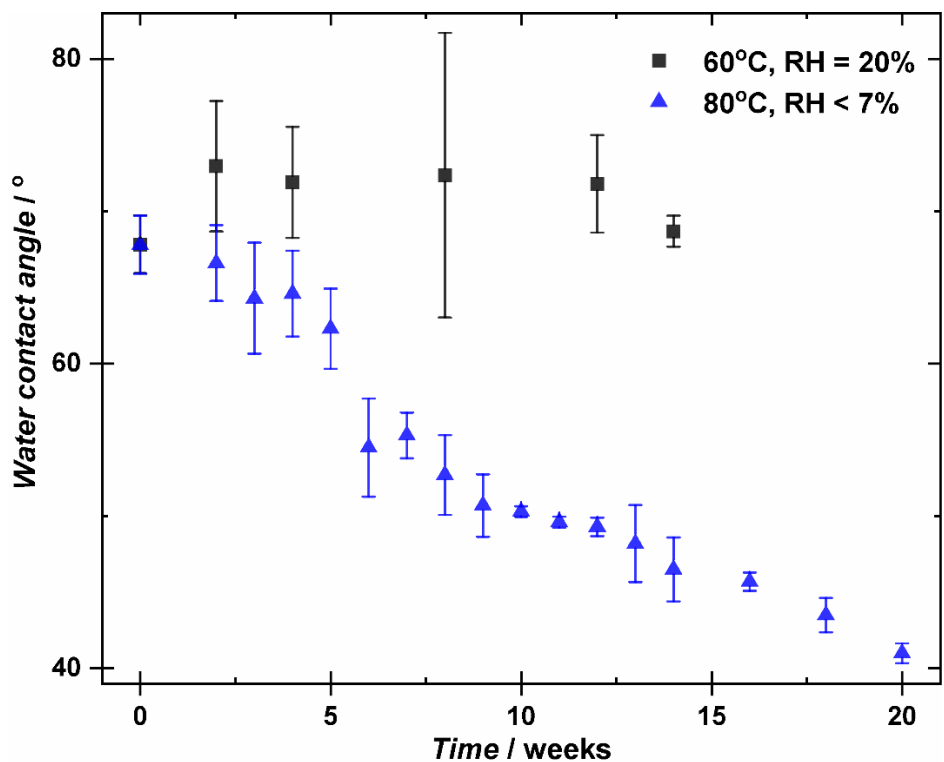

**Figure S11.** Water contact angle measurements for PVC1 degraded at 60°C, RH = 20% and 80°C, without humidity control (Case I).

**Table S4.** The elemental composition of pristine and degraded PVC1 samples measured by CHN elemental analysis.

| CHN+S analysis of undegraded PVC1 sample | | | | | |
|---|---|---|---|---|---|
| Sample | Carbon content /% | Hydrogen content /% | Nitrogen content /% | Sulfur content /% | The remaining mass of a sample /% |
| PVC1 (undegraded) | 41.91 | 5.280 | 0.07 | 0.266 | 52.47 |
| **PVC1 degraded at 60˚C, RH = 60%** | | | | | |
| Degradation time /weeks | Carbon content /% | Hydrogen content /% | Nitrogen content /% | The remaining mass of a sample /% | Change in the remaining mass /% |
| 0 | 41.85 | 5.339 | 0.05 | 52.76 | x |
| 4 | 41.84 | 5.193 | 0.09 | 52.87 | 0.11 |
| 8 | 41.84 | 5.156 | 0.08 | 52.93 | 0.17 |
| 12 | 41.91 | 5.099 | 0.10 | 52.89 | 0.13 |
| 16 | 41.87 | 5.117 | 0.06 | 52.95 | 0.19 |
| 20 | 41.62 | 5.118 | 0.10 | 53.16 | 0.40 |
| **PVC1 degraded at 80˚C, no RH control** | | | | | |
| Degradation time /weeks | Carbon content /% | Hydrogen content /% | Nitrogen content /% | The remaining mass of a sample /% | Change in the remaining mass /% |
| 0 | 41.85 | 5.339 | 0.05 | 52.76 | x |
| 4 | 42.25 | 5.369 | 0.14 | 52.24 | -0.51 |
| 8 | 42.83 | 5.322 | 0.12 | 51.73 | -1.03 |
| 12 | 42.44 | 5.324 | 0.12 | 52.12 | -0.63 |
| 16 | 42.28 | 5.162 | 0.11 | 52.45 | -0.31 |
| 20 | 42.52 | 5.370 | 0.12 | 51.99 | -0.77 |

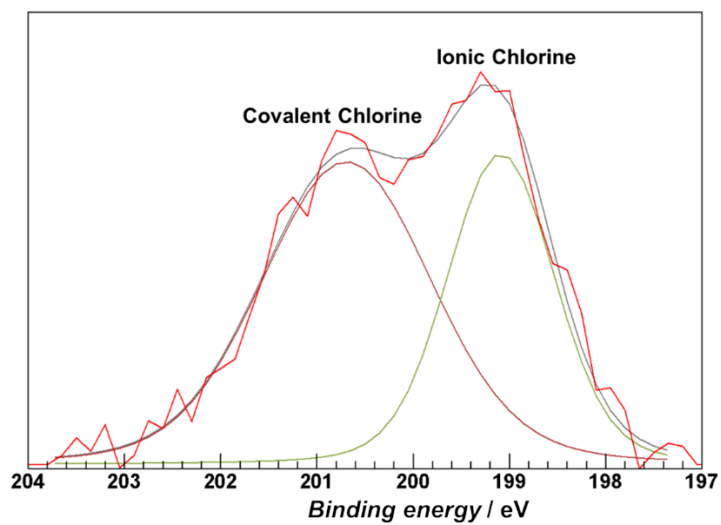

**Figure S12.** The XPS Cl $2p^{3/2}$ and Cl $2p^{1/2}$ spectra of PVC degraded at 80˚C with no control humidity for 22 weeks (example of Case I).

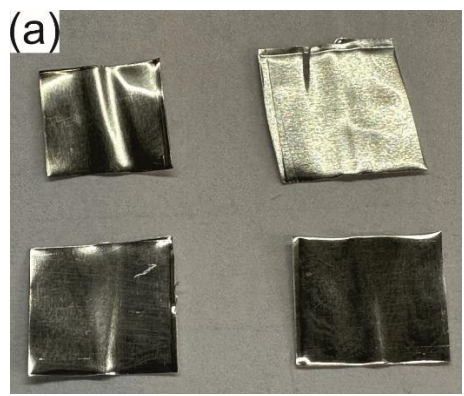

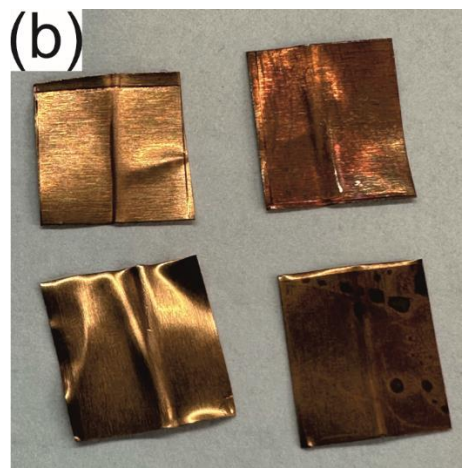

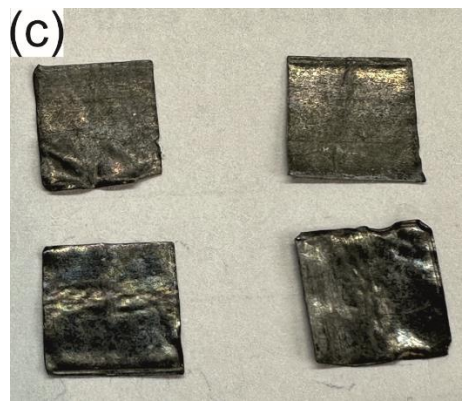

**Figure S13.** Pictures of metal foils after the Oddy test – silver **(a)**, copper **(b)**, and lead **(c)**. On the left side are the control samples (conducted without the PVC1 sample), and on the right side are the actual samples (from the Oddy test conducted with the PVC1 sample).

The Oddy test, as presented in **Figure S13** for PVC1, was proposed in the 1970s as a test to evaluate the safety of materials in and around objects, especially concerning heritage institutions and valuable artifacts. The main purpose is to detect the release of volatile compounds, such as HCl in the case of polyvinyl chloride (PVC), that could damage metals or other materials in enclosed environments. It involves placing a sample, here – 2 g of PVC1, in sealed containers with metal foils (silver **(a)**, copper **(b),** and lead **(c)**) at 60°C and 100% RH, followed by assessing the extent of corrosion after 28 days. The metal foils in **Figure S13** show no significant signs of corrosion or tarnishing; only slight changes are visible on the copper coupons (**Figure S13c**).

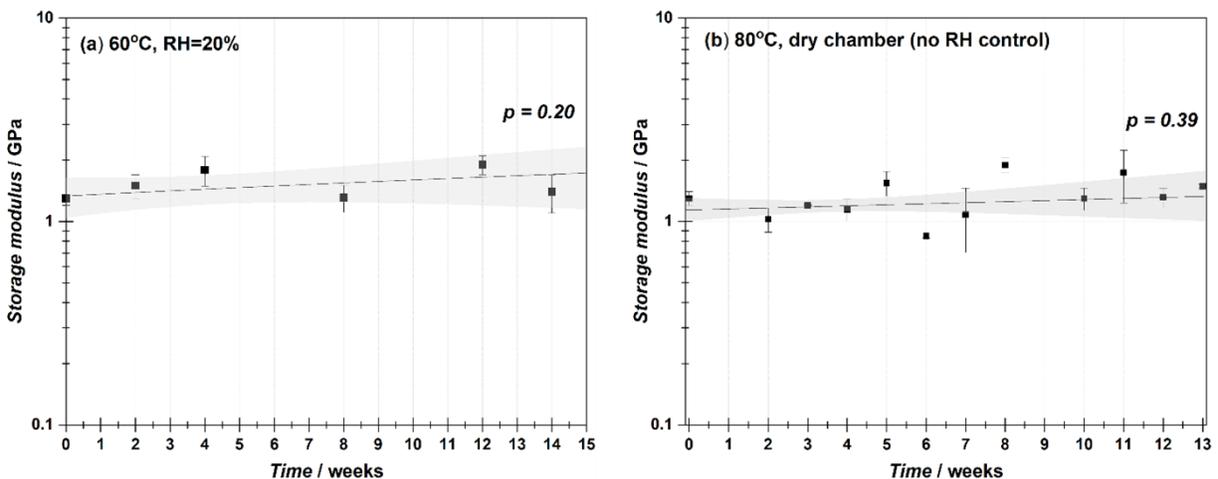

**Figure S14.** Storage modulus $E'$ evolution with degradation time for the PVC1 degraded at low RH (Case I): 60°C, RH = 20% **(a)** and 80°C, without humidity control **(b).** The grey area represents confidence bands corresponding to the confidence level of 95%.

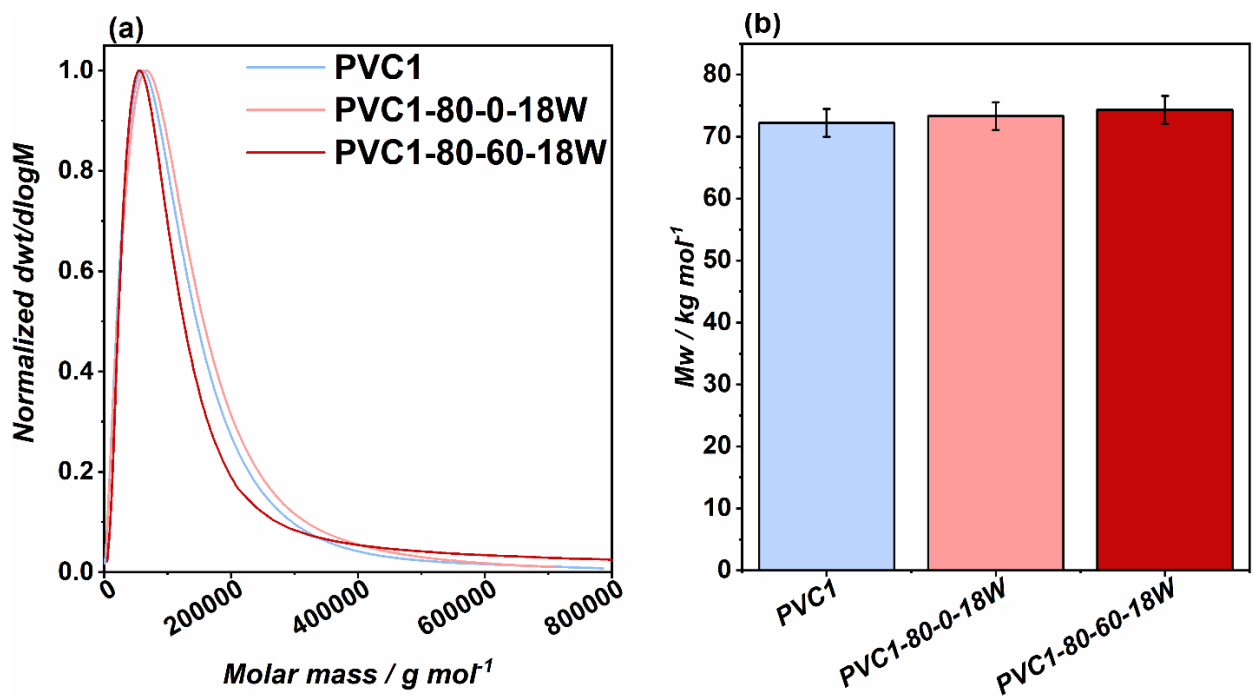

**Figure S15.** SEC results: molecular weight distributions **(a)**, and $M_w$ values **(b)** of undegraded PVC1 and PVC1 degraded at 80°C with no RH control and at RH = 60% after 18 weeks (PVC1-80-0-18W, PVC1-80-60-18W, respectively).